\documentclass[10pt,a4paper]{article}

\usepackage{geometry}
\usepackage{graphicx}

\usepackage{natbib}

\usepackage{amsmath}
\usepackage{booktabs}

\usepackage{amssymb}
\usepackage{subcaption}
\usepackage{relsize}
\usepackage{comment}
\usepackage{listings}
\usepackage{bm}

\usepackage[colorlinks, linkcolor=blue, citecolor=blue, urlcolor  = blue]{hyperref}

\newcommand{\pkg}[1]{\texttt{#1}}

\title{\textbf{matrixdist: An R Package for Statistical Analysis of Matrix Distributions}}

\author{Martin Bladt$^{(1)}$, Alaric Mueller$^{(2)}$ and Jorge Yslas$^{(3)}$}

\date{
{\small
$^{1}$Department of Mathematical Sciences, 
University of Copenhagen,
DK-2100 Copenhagen, 
Denmark \\
$^{2}$Faculty of Business and Economics,
University of Lausanne,
1015 Lausanne,
Switzerland\\
$^{3}$Institute for Financial and Actuarial Mathematics,
University of Liverpool,
L69 7ZL Liverpool,
UK
}
}

\begin{document}

\maketitle

\begin{abstract}
The \pkg{matrixdist} R package provides a comprehensive suite of tools for the statistical analysis of matrix distributions, including phase-type, inhomogeneous phase-type, discrete phase-type, and related multivariate distributions. This paper introduces the package and its key features, including the estimation of these distributions and their extensions through expectation-maximisation algorithms, as well as the implementation of regression through the proportional intensities and mixture-of-experts models. Additionally, the paper provides an overview of the theoretical background, discusses the algorithms and methods implemented in the package, and offers practical examples to illustrate the application of \pkg{matrixdist} in real-world  actuarial problems. The \pkg{matrixdist} R package aims to provide researchers and practitioners a wide set of tools for analysing and modelling complex data using matrix distributions.
\end{abstract}

\tableofcontents

\section{Introduction}
In recent years, matrix distributions such as phase-type (PH), inhomogeneous phase-type (IPH), discrete phase-type (DPH), and related multivariate distributions have gained prominence in various fields due to their ability to seamlessly model complex or heterogeneous data. Matrix distributions can be employed to capture diverse behaviours and dependencies in real-world scenarios, making them particularly useful in areas such as survival analysis, actuarial science, queuing theory, and finance.

PH distributions \citep[cf.][]{bladt2017matrix} are a natural extension of the exponential distribution and are defined as the time to reach an absorbing state in an otherwise transient time-homogeneous pure-jump Markov process. These distributions possess closed-form formulas in terms of matrices for various functionals such as density, cumulative distribution function, Laplace transform, moments, and can approximate any other distribution on the positive real line. Since only the time to absorption is observed and not the actual stochastic process trajectory, models based on PH distributions are hidden Markov models. Hence, they are typically estimated using the expectation-maximisation (EM) algorithm \citep[cf.][]{asmussen1996fitting}. However, their tail behaviour remains exponential, which can be a limiting factor in certain applications that require a more accurate tail description. This shortcoming has led to the exploration of classes extending PH distributions, offering different tail behaviours while retaining PH-like properties. One such class is comprised of IPH distributions \citep{albrecher2019inhomogeneous}, which can be defined equivalently as the law of a transformed PH distributed random variable or the absorption time of a time-inhomogeneous pure-jump Markov process. The tail behaviour of the resulting IPH distribution can be as heavy or light as desired, depending on the chosen transformation. Moreover, for parametric transformations, the estimation procedure has been outlined in \cite{albrecher2020fitting}. Several extensions to the multivariate setting of PH distributions have been introduced in the literature. The most general construction can be found in \cite{kulkarni89}, called the MPH* class, whose definition consists of considering marginals that depend on a unique Markov jump process but with different rewards at each state. Regarding multivariate extensions of the IPH class, \cite{albrecher2020fitting} presented some alternatives that allow for estimation. Other extensions were considered in \cite{albrecher2020fitting} and \cite{bladt2022mph}.

For modelling count data, the DPH class \citep[cf.][]{bladt2017matrix} provides a tractable and flexible set of distributions, which also allows for estimation via an EM algorithm. Formally, DPH distributions are defined as the time to reach absorption in a discrete Markov chain with one absorbing state and the remaining states being transient. However, they can also be viewed as an extension of the geometric distribution. The more general class of multivariate DPH distributions was introduced in \cite{navarro2018order}, employing a similar construction principle as the MPH* class via rewards. More recently, \cite{bladt2023robust} presented other more tractable classes of DPH distributions that offer similar versatility in modelling count data.

Matrix distributions have proven to be useful in various applications, motivating the development of regression models tailored to these distributions. Two notable regression models for matrix distributions include the proportional intensities (PI) model and the mixture-of-experts (MoE) specification. The PI model, as proposed by \cite{albrecher2022mortality}, extends the classical Cox proportional hazards model to the IPH setting. This model enables the incorporation of covariate information directly into the intensity matrix of the underlying Markov process. On the other hand, the MoE specification, originally introduced in \cite{bladtyslas2022}, incorporates the covariate information in the vector of initial probabilities and can easily be modified to the multivariate case \citep[cf.][]{albrecher2022joint}.

This paper presents the \pkg{matrixdist} package \citep{matrixdist}, designed to work with matrix distributions and implemented in R \citep{R}. The \pkg{matrixdist} package leverages the S4 class system in R to ensure a well-structured and organised implementation of the various distribution classes and their associated methods. By utilising S4 classes, the package offers a user-friendly interface for interacting with PH, IPH, DPH, and related multivariate and regression distributions. Users can create distribution objects with their respective parameters, and the package will automatically manage the underlying data structures and computations. This design promotes code reusability, maintainability, and extensibility, making it easier to add new distribution classes and methods in the future while ensuring a consistent and coherent user experience. The object oriented design can be particularly efficient when dealing with large numbers of parameters during statistical estimation. Furthermore, most computationally intensive functions are implemented using the C++ language to enhance performance. These functions are then made accessible in the R language via the \pkg{Rcpp} package, ensuring both speed and ease-of-use for the \pkg{matrixdist} package.  The \pkg{matrixdist} package is now available in CRAN (\url{https://cran.r-project.org/package=matrixdist}) and can be installed through \texttt{install.packages("matrixdist")}. The source code is available in the GitHub repository \url{https://github.com/martinbladt/matrixdist_1.0}, where bug reports can also be submitted.

It is worth mentioning that the R package \pkg{matrixdist} offers a preferable solution compared to existing packages for working with matrix distributions. To provide context, we briefly summarise some alternative software tools for matrix distributions:
\begin{itemize} 
\item The R package \pkg{actuar} \citep{dutang2008actuar} contains implementations for the density, cumulative distribution, moments, and moment-generating function of univariate PH distributions.
\item The R package \pkg{mapfit} \citep{okamura2015mapfit} offers some estimation methods for PH distributions and for Markovian Arrival Processes (MAP). The latter can loosely be seen as dependent concatenations of PH variables
    (arrivals).
\item The recent \pkg{PhaseTypeR} package \citep{phasetypeR} includes implementations for reward transformations and functionals for some of the multivariate extensions of homogeneous PH and DPH distributions. However, no statistical estimation is considered.
\end{itemize}

The \pkg{matrixdist} package stands out by providing a comprehensive and efficient suite of tools that cover univariate, multivariate, continuous, discrete, and regression matrix models, and their statistical estimation. Right-censoring is also supported. The package's versatility, completeness, and high-speed performance make it an ideal choice for researchers and practitioners across all matrix distribution domains.

The remainder of the paper is organised as follows. Section~\ref{sec:uni} provides the mathematical formulation of univariate DPH, PH and IPH distributions, their basic properties and regression extensions. Similar information for multivariate matrix distribution is given in Section~\ref{sec:multi}, where the multivariate extensions of the IPH and DPH classes are introduced. A discussion on estimation methods for the introduced models via EM algorithms is provided in Section~\ref{sec:fit}, and detailed illustrations on the use of the package are provided in Section~\ref{sec:ill}. Finally, Section~\ref{sec:sum} gives a condensed summary of the methods available in \pkg{matrixdist} and shows how further information can be accessed within the package's documentation.

\section{Univariate matrix distributions} \label{sec:uni}

\subsection{Discrete phase-type distributions}

Let $(Z_n)_{n\in \mathbb{N}_0}$ be a discrete Markov chain (MC) on a finite state space $\{1, \dots, p, p+1\}$ where the first $p$ states are transient, and state $p+1$ is absorbing. Then, the MC has transition
matrix $\mathbf{P}$ given by 
$$
    \mathbf{P}= \left( \begin{array}{cc}
        \mathbf{S} &  \mathbf{s} \\
        \boldsymbol{0} & 1
    \end{array} \right)\,, 
$$
where $\mathbf{S}$ is a $p\times p$ matrix, known as sub-transition matrix, and $\mathbf{s}$ is a column vector of dimension $p$. Considering that the rows of $\mathbf{P}$ sum to $1$, the following relation holds $\mathbf{s}=\mathbf{e}-\mathbf{S}\mathbf{e}$, with $\mathbf{e} = (1, \dots, 1)^{\top}$ the $p$-dimensional column vector of ones. Furthermore, we assume that $(Z_n)_{n\in \mathbb{N}_0}$ may start in any transient state with a certain probability $\alpha_k=\mathbb{P}(Z_0=k)$, $k = 1, \dots, p$, and let $\boldsymbol{\alpha} = (\alpha_1 ,\dots,\alpha_p )$. Then, the time until absorption $N=\inf\{n \ge 1 \mid Z_n=p+1 \}$ is said to be discrete phase-type (DPH) distributed with representation $(\boldsymbol{\alpha},\mathbf{S})$, and we write $N \sim \mbox{DPH}(\boldsymbol{\alpha},\mathbf{S})$ or $N \sim \mbox{DPH}_{p}(\boldsymbol{\alpha},\mathbf{S})$. Such a random variable has probability mass function $q$ and distribution function $Q$ given by 
$$
     q(n) =  \boldsymbol{\alpha}\mathbf{S}^{n-1}\mathbf{s} \,, \quad n\ge 1 \,,
$$ 
$$
    Q(n) = 1-\boldsymbol{\alpha}\mathbf{S}^{n}\mathbf{e} \,, \quad n\ge 1 \,.
$$

The absorption time $N$ has $\kappa$-factorial moments, $\kappa \in \mathbb{N}$, given by
$$
\mathbb{E}(N(N - 1)\cdots (N - \kappa + 1))=\kappa! \boldsymbol{\alpha}\mathbf{S}^{\kappa - 1}\left(\mathbf{I}-\mathbf{S}\right)^{-\kappa}\mathbf{e} \,.
$$

In particular, we have that $\mathbb{E}(N)=\boldsymbol{\alpha}\left(\mathbf{I}-\mathbf{S}\right)^{-1}\mathbf{e}$. Furthermore, its probability-generating function $P_N$ is given by
$$
P_N(t) = \mathbb{E}(t^N)= \boldsymbol{\alpha} \left( t^{-1}\mathbf{I} - \mathbf{S}  \right)\mathbf{s} \,, \quad |t| \leq 1 \, ,
$$
where $\mathbf{I}$ is the identity matrix of appropriate dimension.

Let $N_1 \sim \mbox{DPH}_{p_1}(\boldsymbol{\alpha}_1,\mathbf{S}_1)$ and $N_2 \sim \mbox{DPH}_{p_2}(\boldsymbol{\alpha}_2,\mathbf{S}_2)$ be two independent DPH distributed random variables. Then, 
$$
    N_1 + N_2 \sim \mbox{DPH}_{p_1 + p_2} \left( \left( \boldsymbol{\alpha}_1, \boldsymbol{0}\right), \left(\begin{array}{cc}
        \mathbf{S}_1 &  \mathbf{s}_1 \boldsymbol{\alpha}_2 \\
        \boldsymbol{0} & \mathbf{S}_2
    \end{array} \right) \right)\,,
$$
$$
    \min\left(N_1, N_2 \right) \sim \mbox{DPH}_{p_1 p_2}\left( \boldsymbol{\alpha}_1 \otimes  \boldsymbol{\alpha}_2, 
        \mathbf{S}_1 \oplus \mathbf{S}_2  \right)\,, 
$$
and 
$$
    \max\left(N_1, N_2 \right) \sim \mbox{DPH}_{p_1 p_2 + p_1 + p_2}\left( \left( \boldsymbol{\alpha}_1 \otimes  \boldsymbol{\alpha}_2, \boldsymbol{0}, \boldsymbol{0}\right), \left(\begin{array}{ccc}
        \mathbf{S}_1 \otimes \mathbf{S}_2  &  \mathbf{S}_1 \otimes \mathbf{s}_2 & \mathbf{s}_1 \otimes \mathbf{S}_2  \\
        \boldsymbol{0} & \mathbf{S}_1 & \boldsymbol{0} \\
        \boldsymbol{0}  & \boldsymbol{0} & \mathbf{S}_2
    \end{array} \right) \right)\,,
$$
where $\otimes$ and $\oplus$ denote the Kronecker product and sum, respectively. Furthermore, if $U \sim Bernoulli(\nu)$, $\nu \in [0, 1]$, then 
$$
    U N_1 + (1 - U) N_2  \sim \mbox{DPH}_{p_1 + p_2}\left( \left(
        \nu \boldsymbol{\alpha}_1 \,,  (1 - \nu) \boldsymbol{\alpha}_2  \right), 
      \left( \begin{array}{cc}
        \mathbf{S}_1  &  \boldsymbol{0} \\
        \boldsymbol{0} & \mathbf{S}_2
    \end{array} \right) \right)\,.
$$

Thus, the DPH class is closed under addition, minima, maxima, and finite mixtures. We refer to \cite{bladt2017matrix} for a detailed description of DPH distributions.

\subsubsection*{Regression} 

Recently, a regression model for frequency modelling based on DPH distributions was introduced in \cite{bladt2023robust} as a generalisation of mixture-of-experts (MoE) models. The main idea is to incorporate the covariate information on the vector of initial probabilities, which play the role of expert weights. The resulting distributional model is flexible, as it possesses the property of denseness on more general regression models satisfying some mild technical conditions. Next, we present the fundamental concepts of this regression model. Define the mapping 
$$
\boldsymbol{\alpha}:D \subset \mathbb{R}^h \rightarrow \Delta^{p-1}, 
$$ 
where $\Delta^{p-1}=\{(\alpha_1,\dots,\alpha_p)\in \mathbb{R}^p\mid \sum_k \alpha_k=1,\; \alpha_k\ge 0 \; \forall k \}$ is the standard $(p-1)$ simplex. Then, for any given vector of covariate information $\mathbf{X} = (X_1, \dots, X_h) \in \mathbb{R}^h$, the initial probabilities of the underlying MC of a DPH distribution are taken to be $\mathbb{P}(Z_0=k)=\alpha_k(\mathbf{x}):=(\boldsymbol{\alpha}(\mathbf{x}))_k$, $k=1,\dots,p$. Thus, we say that $N\mid \mathbf{X}\sim\mbox{DPH}(\boldsymbol{\alpha}(\mathbf{X}),\mathbf{S})$ follows the DPH-MoE specification. For $D=\mathbb{R}^h$, we have a convenient parametrisation of initial probabilities, namely the softmax parametrisation given by 
$$
    \alpha_k(\mathbf{X};\boldsymbol{\gamma})=\frac{\exp\left(\mathbf{X}\boldsymbol{\gamma}_k\right)}{\sum_{j=1}^{p}\exp\left(\mathbf{X}\boldsymbol{\gamma}_j\right)} \quad k=1,\dots,p\;,
$$
where $\boldsymbol{\gamma}_k\in \bar{\mathbb{R}}^h$, and $\boldsymbol{\gamma}=(\boldsymbol{\gamma}_1^{\top},\dots,\boldsymbol{\gamma}_p^{\top})^{\top}\in\bar{\mathbb{R}}^{p\times h}$. More details about this model as well as precise denseness considerations can be found in \cite{bladt2023robust}. We also refer to \cite{bladtyslas2022} for an extension of this regression model to continuous phase-type distributions.

\subsection{Continuous phase-type distributions}

Let $( J_t )_{t \geq 0}$ be a time-homogeneous Markov jump process on a finite state space $\{1, \dots, p, p+1\}$, where states $1,\dots,p$ are transient, and state $p+1$ is absorbing. Then, the intensity matrix $\boldsymbol{\Lambda}$ of $( J_t )_{t \geq 0}$ is of the form 
$$
    \boldsymbol{\Lambda}= \left( \begin{array}{cc}
        \mathbf{S} &  \mathbf{s} \\
        \boldsymbol{0} & 0
    \end{array} \right)\,, 
$$ 
where $\mathbf{S}$ is a $p \times p$ matrix, called a sub-intensity matrix, and $\mathbf{s}$ is a $p$-dimensional column vector. Since every row of $\boldsymbol{\Lambda}$ sums to zero, it follows that $\mathbf{s}=- \mathbf{S} \, \mathbf{e}$. Assume that the process starts somewhere in the transient space with probability $\alpha_{k} = \mathbb{P}(J_0 = k)$, $k = 1,\dots, p$, and let $\boldsymbol{\alpha} = (\alpha_1 ,\dots,\alpha_p )$. Here it is assumed that the probability of starting in the absorbing state $p+1$ is zero, i.e., $\mathbb{P}(J_0 = p + 1) = 0$. Then, we say that the time until absorption $Y$ has phase-type (PH) distribution with representation $(\boldsymbol{\alpha},\mathbf{S} )$, and we write $Y \sim \mbox{PH}(\boldsymbol{\alpha},\mathbf{S} )$ or $Y \sim \mbox{PH}_{p}(\boldsymbol{\alpha},\mathbf{S} )$. The density $f$ and distribution function $F$ of $Y \sim \mbox{PH}(\boldsymbol{\alpha},\mathbf{S} )$ are given by
$$
     f(y) = \boldsymbol{\alpha} \exp\left(\mathbf{S} y\right) \mathbf{s} \,, \quad y>0 \,,
$$
$$
    F(y) = 1-\boldsymbol{\alpha} \exp\left(\mathbf{S} y\right) \mathbf{e} \,, \quad y>0 \,,
$$
where the exponential of a matrix $\mathbf{A}$ is defined by 
$$
    \exp\left(\mathbf{A} \right) = \sum_{n=0}^{\infty} \frac{\mathbf{A}^{n}}{n!} \,.
$$ 

The efficient computation of the exponential of a matrix is not a straightforward task in high dimensions, and we refer to \cite{moler1978nineteen} for a survey of different methods. The moments of $Y \sim \mbox{PH}(\boldsymbol{\alpha},\mathbf{S} )$ are given by
$$
  \mathbb{E}\left( Y^{\theta} \right) = \Gamma \left( 1+ \theta\right) \boldsymbol{\alpha} \left(- \mathbf{S} \right)^{-\theta} \mathbf{e}\,, \quad \theta > 0 \,.
$$

Moreover, its Laplace transform $L_Y$ is given by
$$
  L_Y(u) = \mathbb{E}\left( \exp(-u Y) \right) =  \boldsymbol{\alpha} \left( u \mathbf{I} - \mathbf{S} \right) \mathbf{s} \,, \quad u \geq 0 \,.
$$

Let $Y_1 \sim \mbox{PH}_{p_1}(\boldsymbol{\alpha}_1,\mathbf{S}_1)$ and $Y_2 \sim \mbox{PH}_{p_2}(\boldsymbol{\alpha}_2,\mathbf{S}_2)$ be independent PH distributed random variables. Then 
$$
    Y_1 + Y_2 \sim \mbox{PH}_{p_1 + p_2} \left( \left( \boldsymbol{\alpha}_1, \boldsymbol{0}\right), \left(\begin{array}{cc}
        \mathbf{S}_1 &  \mathbf{s}_1 \boldsymbol{\alpha}_2 \\
        \boldsymbol{0} & \mathbf{S}_2
    \end{array} \right) \right)\,,
$$ 
$$
    \min\left(Y_1, Y_2 \right) \sim \mbox{PH}_{p_1 p_2}\left( \boldsymbol{\alpha}_1 \otimes  \boldsymbol{\alpha}_2, 
        \mathbf{S}_1 \oplus \mathbf{S}_2  \right)\,, 
$$
and 
$$
    \max\left(Y_1, Y_2 \right) \sim \mbox{PH}_{p_1 p_2 + p_1 + p_2}\left( \left( \boldsymbol{\alpha}_1 \otimes  \boldsymbol{\alpha}_2, \boldsymbol{0}, \boldsymbol{0}\right), \left(\begin{array}{ccc}
        \mathbf{S}_1 \oplus \mathbf{S}_2  &  \mathbf{I} \otimes \mathbf{s}_2 & \mathbf{s}_1 \otimes \mathbf{I}  \\
        \boldsymbol{0} & \mathbf{S}_1 & \boldsymbol{0} \\
        \boldsymbol{0}  & \boldsymbol{0} & \mathbf{S}_2
    \end{array} \right) \right)\,,
$$

This means that the PH class is closed under addition, minima, and maxima. In fact, it is also closed under any order statistic (cf. \cite{bladt2017matrix} for the exact and more involved expression). Moreover, the class is closed under finite mixtures. Concretely, if $U \sim Bernoulli(\nu)$, $\nu \in [0, 1]$, then 
$$
    U Y_1 + (1 - U) Y_2  \sim \mbox{PH}_{p_1 + p_2}\left( \left(
        \nu \boldsymbol{\alpha}_1 ,   (1 - \nu) \boldsymbol{\alpha}_2 
    \right), 
       \left( \begin{array}{cc}
        \mathbf{S}_1  &  \boldsymbol{0} \\
        \boldsymbol{0} & \mathbf{S}_2
    \end{array} \right) \right)\,. 
$$

Furthermore, when the mixing probabilities are given by a DPH distribution, the resulting distribution is again PH. More specifically, let $N \sim \mbox{DPH}_{q}(\boldsymbol{\pi},\mathbf{T})$ and $Y_1, Y_2 \dots$ be iid, independent of $N$, with common element $Y \sim \mbox{PH}_{p}(\boldsymbol{\alpha},\mathbf{S})$. Then 
$$
\sum_{n = 1}^{N} Y_n \sim \mbox{PH}_{qp}(\boldsymbol{\pi}\otimes\boldsymbol{\alpha},\mathbf{I}\otimes\mathbf{S} + \mathbf{T}\otimes \mathbf{s}\boldsymbol{\alpha}) \,.
$$ 

We refer to \cite{bladt2017matrix} for a comprehensive reading on PH distributions.

By decomposing the matrix $\mathbf{S}$ into its Jordan normal form, we observe that the right tail of a PH distribution exhibits asymptotically exponential behaviour. This limitation motivates the introduction of PH extensions that can accommodate diverse tail behaviours. One such extension is the class of inhomogeneous phase-type distributions, which is presented in the following section.

\subsection{Inhomogeneous phase-type distributions}

In \cite{albrecher2019inhomogeneous}, the class of inhomogeneous phase-type distribution was introduced by allowing the Markov jump process to be time-inhomogeneous in the construction principle of PH distributions. In this way, $(J_t)_{t \geq 0}$ has an intensity matrix of the form 
$$
    \boldsymbol{\Lambda}(t)= \left( \begin{array}{cc}
        \mathbf{S}(t) &  \mathbf{s}(t) \\
        \boldsymbol{0} & 0
    \end{array} \right)\,, 
$$ 
where $\mathbf{s}(t)=- \mathbf{S}(t) \, \mathbf{e}$. Here it is assumed that the vector of initial probabilities is $\boldsymbol{\alpha} = (\alpha_1 ,\dots,\alpha_p )$. Then, we say that the time until absorption $Y$ has inhomogeneous phase-type (IPH) distribution with representation $(\boldsymbol{\alpha},\mathbf{S}(t) )$, and we write $Y \sim \mbox{IPH}(\boldsymbol{\alpha},\mathbf{S}(t) )$. However, this setting is too general for applications since its functionals are given in terms of product integrals. Thus, we consider the subclass of IPH distribution when $\mathbf{S}(t)$ is of the form $\mathbf{S}(t) =\lambda(t)\mathbf{S}$, where $\lambda$ is some known non-negative real function, and $\mathbf{S}$ is a sub-intensity matrix. In this case, we write $Y \sim \mbox{IPH}(\boldsymbol{\alpha},\mathbf{S}, \lambda )$. Note that with $\lambda \equiv 1$, one gets the conventional PH distributions. This subclass is particularly suitable for applications due to the following property. If $Y \sim \mbox{IPH}(\boldsymbol{\alpha},\mathbf{S}, \lambda )$, then there exists a function $g$ such that 
$$
    Y \sim g(\tau) \,,
$$ 
where $\tau \sim \mbox{PH}(\boldsymbol{\alpha},\mathbf{S})$. The relationship between $g$ and $\lambda$ is given by the following expression 
$$
    g^{-1}(y) = \int_{0}^{y} \lambda(t) dt   \,.
$$ 

The density $f$ and distribution function $F$ of $Y \sim \mbox{IPH}(\boldsymbol{\alpha},\mathbf{S}, \lambda )$ can again be expressed using the exponential of matrices and are given by
$$
  f(y) = \lambda (y)\, \boldsymbol{\alpha}\exp \left( \int_0^y \lambda (t) dt\ \mathbf{S} \right)\mathbf{s}\,,\quad y>0 \,,
$$ 
$$
   F(y) = 1- \boldsymbol{\alpha} \exp \left( \int_0^y \lambda (t)dt\ \mathbf{S} \right) \mathbf{e}\,, \quad y>0 \,.
$$

The class of IPH distributions is no longer confined to exponential tails, despite being defined in terms of matrix exponentials. Instead, the function $\lambda$ determines their exact asymptotic behaviour (see \cite{albrecher2022mortality} for details).

It turns out that a number of IPH distributions can be expressed as classical distributions with matrix-valued parameters properly defined using functional calculus. Table \ref{tab:trans} contains the IPH transformations as implemented in the \pkg{matrixdist} package (see \cite{albrecher2019inhomogeneous, albrecher2020fitting,
albrecher2022mortality} for further information and rationale on the different parametrisations). 

\begin{table}[h]
\centering
\begin{tabular}{lccc}
  \hline
                    &   $\lambda(t)$       & $g(y)$ & Parameters Domain  \\
  \hline
Matrix-Pareto       & $( t + \beta )^{-1}$ & $ \beta \left( \exp(y)-1\right)$ & $\beta>0$ \\[2mm] 
Matrix-Weibull      & $\beta t ^{\beta -1}$    & $y^{1/\beta}$ & $\beta>0$ \\ [2mm]
Matrix-Lognormal    & $\gamma (\log(t+1))^{\gamma-1}/(t+1) $    & $\exp(y^{1/\gamma})-1$      & $\gamma > 1$     \\ [2mm]
Matrix-Loglogistic  & $\theta t^{\theta -1} / (t^\theta + \gamma^{\theta})$    & $\gamma (\exp(y)-1)^{1/\theta}$    & $\gamma,\theta >0$     \\ [2mm]
Matrix-Gompertz     & $\exp (\beta t)$    & $\log( \beta y  + 1 ) / \beta$      & $\beta > 0$     \\ [2mm] 
Matrix-GEV          &  -   & $\mu + {\sigma}(y^{-\xi} - 1)/ {\xi}$      & $\mu \in \mathbb{R}$, $\sigma >0$,  $\xi \in \mathbb{R}$    \\[2mm]
   \hline
\end{tabular}
\caption{Transformations}
\label{tab:trans}
\end{table}

\subsubsection*{Regression}

At least two regression models based on IPH distributions exist in the literature. The first model, known as the proportional intensities (PI) model, was introduced in \cite{albrecher2022mortality}. The authors, inspired by the proportional hazards model \citep[cf.][]{cox72reg}, formulated a way of regressing on the intensity function $\lambda$ for different covariate information. More specifically, the main idea of the PI model is that the inhomogeneous intensity $\lambda(\cdot;\boldsymbol{\theta})$, which is assumed to be fully determined by a vector of parameters $\boldsymbol{\theta}$, is proportionally affected for different covariate information $\mathbf{X}\in \mathbb{R}^h$ according to a positive real-valued and measurable function $m(\cdot)$. In other words, the model assumes that 
$$
  \lambda(t ; \mathbf{X},\boldsymbol{\theta}, \boldsymbol{\gamma})=\lambda(t;\boldsymbol{\theta})m(\mathbf{X}\boldsymbol{\gamma}), \quad t\ge 0,
$$
where $\boldsymbol{\gamma}$ is a $h$-dimensional column vector containing regression coefficients. With this assumption, $Y \mid \mathbf{X}\sim\mbox{IPH}(\boldsymbol{\alpha},\mathbf{S}, \lambda(\cdot ; \mathbf{X},\boldsymbol{\theta}, \boldsymbol{\gamma}))$ with corresponding density and cumulative functions 
$$
f(y)= m(\mathbf{X}\boldsymbol{\gamma}) \lambda(y;\boldsymbol{\theta}) \boldsymbol{\alpha}\exp\left(m(\mathbf{X}\boldsymbol{\gamma}) \int_0^y\lambda(t;\boldsymbol{\theta})\,dt \mathbf{S}\right)\mathbf{s}\,,
$$
and 
$$
F(y)=1 - \boldsymbol{\alpha}\exp\left(m(\mathbf{X}\boldsymbol{\gamma}) \int_0^y\lambda(t;\boldsymbol{\theta})\,dt \mathbf{S}\right)\mathbf{e}\,.
$$

Typically, the measurable function $m()$ is taken to be $m(x) = \exp(x)$, which is the default implementation in \pkg{matrixdist}. 
In this framework, when $p = 1$, one recovers the proportional hazard model. However, for $p>1$, the implied hazard functions between different subgroups can deviate from proportionality in the distribution body but are asymptotically proportional in the tail. For more details on the PI model, we refer the reader to \cite{albrecher2022mortality} and \cite{bladt2022phase}.

The second regression model, called the phase-type mixture-of-experts (PH-MoE) model, was introduced in \cite{bladtyslas2022} as a generalisation of mixture-of-experts models to the PH distributions framework and follows the same rationale as the DPH-MoE specification described above (cf. \cite{bladtyslas2022} for further details).

\section{ Multivariate matrix distributions} \label{sec:multi}

In this section, we present the mathematical foundations of the
different classes of multivariate matrix distributions implemented in
\pkg{matrixdist}.

\subsection{ Multivariate discrete phase-type distributions}

\subsubsection*{MDPH* class} 

Let $N\sim\mbox{DPH}_p(\boldsymbol{\alpha},\mathbf{S})$ be a DPH
distributed random variable with underlying Markov chain
$(Z_n)_{n \in \mathbb{N}_0}$. Let
$\mathbf{r}_j=(r_j(1),\dots,r_j(p))^{\top}$ be column vectors of
dimension $p$ taking values in $\mathbb{N}_0^{p}$, $j=1,\dots,d$, and
let $\mathbf{R}=(\mathbf{r}_1,\dots,\mathbf{r}_d)$ be a $p \times d$
matrix, called the reward matrix. We define 
$$
N^{(j)}=\sum_{n = 0}^N r_j(Z_n) \,,
$$ 
for all $j = 1, \dots, d$. Then, we say that the random vector
$\mathbf{N}=(N^{(1)},\dots,N^{(d)})$ has a multivariate discrete
phase-type distribution of the MDPH* type, and we write
$\mathbf{N}\sim\mbox{MDPH*}(\boldsymbol{\alpha},\mathbf{S},\mathbf{R})$.
This class of distributions was introduced in \cite{navarro2018order}, and the
construction principle can be seen as an extension of the so-called
transformation via rewards for univariate DPH distributions. In this
contribution, the author showed that elements in this class possess
explicit expressions for joint probability-generating function, joint
moment-generating function, and joint moments. Moreover, the class is
dense in the class of distributions with support in $\mathbb{N}^d$, and
margins are DPH distributed. However, general closed-form expressions
for the joint density and distribution functions are not available, and
the estimation of this general class is still an open question, limiting
their use in practice. Nonetheless, there are subclasses of MDPH*
distributions with explicit expressions of these two functionals that
preserve the denseness property of the MDPH* class. The
\pkg{matrixdist} package provides implementations for two of these
subclasses, which we describe next.

\subsubsection*{fMDPH class}

Let
$\mathbf{N} = (N^{(1)}, N^{(2)})\sim\mbox{MDPH*}(\boldsymbol{\alpha},\mathbf{S},\mathbf{R})$
with 
$$
\boldsymbol{\alpha}=\left( \boldsymbol{\pi},\boldsymbol{0}\right)\,,\quad 
\mathbf{S}=\left(\begin{array}{cc}
\mathbf{S}_{11} & \mathbf{S}_{12}\\
\boldsymbol{0} &\mathbf{S}_{22}
\end{array}
\right)\,, 
 \quad \text{and} \quad
\mathbf{R}=\left(\begin{array}{cc}
\mathbf{e} & \boldsymbol{0}\\
\boldsymbol{0} &\mathbf{e}
\end{array}\right)\,.
$$

Here, $\mathbf{S}_{11}$ and $\mathbf{S}_{22}$ are sub-transition
matrices of dimensions $p_1$ and $p_2$ ($p_1+p_2=p$), respectively,
$\mathbf{S}_{12}$ is a $p_1 \times p_2$ matrix satisfying
$\mathbf{S}_{11}\mathbf{e}+\mathbf{S}_{12}\mathbf{e}=\mathbf{e}$, and $\boldsymbol{\pi}$ is a $p_1$ dimensional vector of initial probabilities. Then,
one can show that the joint density of this random vector is given by 
$$
f_{\mathbf{N}}(n^{(1)},n^{(2)})=\boldsymbol{\pi}\mathbf{S}_{11}^{n^{(1)} - 1}\mathbf{S}_{12}\mathbf{S}_{22}^{n^{(2)} - 1}\mathbf{s}_2 \,,
$$
 where $\mathbf{s}_2= \mathbf{e}-\mathbf{S}_{22} \mathbf{e}$. In this
case, we say that $\mathbf{N}$ is bivariate discrete phase-type
distributed of the feed-forward type, and we write
$\mathbf{N} \sim\mbox{fMDPH}(\boldsymbol{\pi},\mathbf{S}_{11},\mathbf{S}_{12}, \mathbf{S}_{22})$.
In particular, the marginals are parametrised by
$N^{(1)}\sim\mbox{DPH}(\boldsymbol{\pi},\mathbf{S}_{11})$ and
$N^{(2)}\sim\mbox{DPH}(\boldsymbol{\pi}\left(\mathbf{I}-\mathbf{S}_{11}\right)^{-1}\mathbf{S}_{12},\mathbf{S}_{22})$.
For more details on this class of bivariate DPH distributions, we refer
the reader to \cite{bladt2023robust}. Although the above construction
principle can be extended to higher dimensions, the implementation of
the methods associated with this class becomes more challenging.
Fortunately, another subclass of MDPH* distributions introduced in
\cite{bladt2023robust} can easily be tracked and implemented in higher
dimensions, as we describe next.

\subsubsection*{mDPH class} 

Let $(Z_n^{(j)})_{n\in \mathbb{N}_0}$, $j=1,\dots,d$, be Markov chains
on a common state space with $p$ transient states. All chains are
assumed to start in the same state and then evolve independently until
respective absorptions. Formally, 
$$
Z_0^{(j)}=Z_{0}^{(l)}\,,\quad (Z_n^{(j)})_{n\in \mathbb{N}_0} {\perp\!\!\!\!\perp}_{Z_0^{(1)}} (Z_n^{(l)})_{n\in \mathbb{N}_0,\:l \neq j}\,, \quad \forall j,l \in \{1,\dots,d\}\,.
$$

If $N^{(j)}$, $j=1,\dots,d$, are the univariate DPH distributed
absorption times of $(Z_n^{(j)})_{n\in \mathbb{N}_0}$, we say that the
vector $\mathbf{N}=( N^{(1)},\ldots, N^{(d)} )$ has a multivariate
discrete phase-type distribution of the mDPH type, and we write $$
\mathbf{N}\sim \mbox{mDPH}(\boldsymbol{\alpha},\widetilde{S}) \,,
$$ where $\boldsymbol{\alpha}$ is the common vector of initial
probabilities and $\widetilde{S}=\{\mathbf{S}_1,\dots,\mathbf{S}_d\}$, with
$\mathbf{S}_j$ the sub-transition matrix associated with
$(Z_n^{(j)})_{n\in \mathbb{N}_0}$, $j=1,\dots,d$. Explicit expressions
for several functionals of interest of this class can easily be obtained
by conditioning on the starting value of the Markov chains. For
instance, the joint density function of
$\mathbf{N}\sim \mbox{mDPH}(\boldsymbol{\alpha},\widetilde{S})$ is given by
$$
  f_\mathbf{N}(\mathbf{n})=\sum_{k=1}^p \alpha_k \prod_{j=1}^d\mathbf{e}_k^{\top}\mathbf{S}_j^{n^{(j)} - 1} \mathbf{s}_j \,, \quad \mathbf{n}\in\mathbb{N}^d \,.
$$

\subsection{Multivariate continuous phase-type distributions}

We now present the MPH*, fMPH, and mPH classes of distributions, which
follow similar construction principles as their discrete counterparts.

Let $\tau\sim\mbox{PH}_p(\boldsymbol{\alpha},\mathbf{S})$ be a PH
distributed random variable, $\mathbf{r}_j=(r_j(1),\dots,r_j(p))^{\top}$
be nonnegative column vectors of dimension $p$, $j=1,\dots,d$, and
$\mathbf{R}=(\mathbf{r}_1,\dots,\mathbf{r}_d)$ be a $p \times d$ reward
matrix. Vectors $\mathbf{r}_j$ represent rates at which a reward is
accumulated when $(J_t)_{t\ge0}$ is in a specific state. Thus, we denote
by $$
Y^{(j)}=\int_0^\tau r_j(J_t)\,dt \,,
$$ the total reward earned prior to absorption by each component $j$,
$j = 1, \dots, d$. Then, we say that the random vector
$\mathbf{Y}=(Y^{(1)},\dots,Y^{(d)})$ has a multivariate phase-type
distribution of the MPH* type, and we write
$\mathbf{Y}\sim\mbox{MPH*}(\boldsymbol{\alpha},\mathbf{S},\mathbf{R})$
(see \cite{kulkarni89, bladt2017matrix, albrecher2020fitting} for more
details). One attractive characteristic of this class of multivariate
distributions is that, given a nonnegative vector
$\mathbf{w} = (w_1, \dots, w_d)$,
$<\mathbf{Y}, \mathbf{w}> = \sum_{j = 1}^d w_j Y^{(j)}$ is PH
distributed, with representation
$(\boldsymbol{\alpha}_{\mathbf{w}},\mathbf{S}_{\mathbf{w}})$ say. To
obtain the exact parametrisation, we split the state space
$\{1, \dots, p\} = E_{+} \cup E_{0}$, such that $k \in E_{+}$ if
$(\mathbf{R} \mathbf{w}^{\top})_k >0$ and $k \in E_{0}$ if
$(\mathbf{R} \mathbf{w}^{\top})_k =0$. Then, by a reordering of the
states, we can assume that $\boldsymbol{\alpha}$ and $\mathbf{S}$ can be
written as 
$$
\boldsymbol{\alpha}=\left(\begin{array}{cc}
        \boldsymbol{\alpha}^{+}\,, & \boldsymbol{\alpha}^{0}
    \end{array}\right) \quad \text{and} \quad 
    \mathbf{S}=\left(\begin{array}{cc}
        \mathbf{S}^{++} & \mathbf{S}^{+0} \\
        \mathbf{S}^{0+} & \mathbf{S}^{00}
    \end{array}\right) \,,
$$
 where the $+$ terms correspond to states in $E_{+}$, and the $0$
terms to those states in $E_{0}$. For instance, $\mathbf{S}^{0+}$
collects transition intensities by which $(J_t)_{t\ge0}$ jumps from
states in $E_{0}$ to states in $E_{+}$. After this arrangement, we can
express the distribution of the continuous part of $<\mathbf{Y}, \mathbf{w}>$ via a PH representation of the form
$$
\boldsymbol{\alpha}_{\mathbf{w}}=\boldsymbol{\alpha}^{+}+\boldsymbol{\alpha}^{0}\left(-\mathbf{S}^{00}\right)^{-1}\mathbf{S}^{0+}
\quad \text{and} \quad  
\mathbf{S}_{\mathbf{w}}=\boldsymbol{\Delta}((\mathbf{R} \mathbf{w}^{\top})_+)^{-1}\left(\mathbf{S}^{++}+\mathbf{S}^{+0}\left(-\mathbf{S}^{00}\right)^{-1}\mathbf{S}^{0+}\right) \,.
$$ 
Here, $\boldsymbol{\Delta}((\mathbf{R} \mathbf{w}^{\top})_+)$ is a
diagonal matrix with entries the vector
$(\mathbf{R} \mathbf{w}^{\top})_+$ formed of the appropriate ordered
values satisfying $(\mathbf{R} \mathbf{w}^{\top})_k >0$. Note that an
atom at zero of size
$\boldsymbol{\alpha}^0\left(\mathbf{I}-\left(-\mathbf{S}^{00}\right)^{-1}\mathbf{S}^{0+}\right)\mathbf{e}$
may appear since it is possible that $(J_t)_{t\ge0}$ starts in a state
in $E_0$ and gets absorbed before reaching a state in $E_+$. In
particular, the above result implies that if
$\mathbf{Y}\sim\mbox{MPH*}(\boldsymbol{\alpha},\mathbf{S},\mathbf{R})$,
then its marginals $Y^{(j)}$ are PH distributed with parameters easily
computed with the aforementioned formulas. Moreover, the so-called
transformation via rewards for univariate PH distributions is retrieved
when $d = 1$ above.

Although members of the MPH* class possess other desirable
characteristics, such as explicit expressions for joint Laplace
transform and joint moments (see Section 8.1.1 of \cite{bladt2017matrix}),
general closed-form expressions for both joint density and joint
distribution functions do not exist. Similar to its discrete
counterpart, the MDPH* class, this complicates their use in practice
and, more importantly, their estimation. This motivates the introduction
of other subclasses of MPH* distributions for which explicit
expressions can be achieved. The \pkg{matrixdist} package provides
implementations for the fMPH and mPH subclasses, which are derived
similarly to their discrete equivalents.

The fMPH class is obtained by considering MPH* parametrisations of the
form $\mbox{MPH*}(\boldsymbol{\alpha},\mathbf{S},\mathbf{R})$ with 
$$
\boldsymbol{\alpha}=\left( \boldsymbol{\pi},\boldsymbol{0}\right)\,,\quad 
\mathbf{S}=\left(\begin{array}{cc}
\mathbf{S}_{11} & \mathbf{S}_{12}\\
\boldsymbol{0} &\mathbf{S}_{22}
\end{array}
\right)\,, 
 \quad \text{and} \quad
\mathbf{R}=\left(\begin{array}{cc}
\mathbf{e} & \boldsymbol{0}\\
\boldsymbol{0} &\mathbf{e}
\end{array}\right)\,,
$$ 
where $\mathbf{S}_{11}$ and $\mathbf{S}_{22}$ are sub-intensity
matrices of dimensions $p_1$ and $p_2$ ($p_1+p_2=p$), respectively,
$\mathbf{S}_{12}$ is a $p_1 \times p_2$ matrix satisfying
$\mathbf{S}_{11}\mathbf{e}+\mathbf{S}_{12}\mathbf{e}=\boldsymbol{0}$, and $\boldsymbol{\pi}$ is a $p_1$ dimensional vector of initial probabilities. In such a case, explicit expressions for different functionals can be
obtained. For example, the joint density of a random vector $\mathbf{Y} = (Y^{(1)}, Y^{(2)})$ with the
above parametrisation is given by 
$$
f_{\mathbf{Y}}(y^{(1)},y^{(2)})=\boldsymbol{\alpha}\exp\left(\mathbf{S}_{11}y^{(1)}\right)\mathbf{S}_{12}\exp\left(\mathbf{S}_{22}y^{(2)}\right)\left(-\mathbf{S}_{22}\right)\mathbf{e} \,,
$$ 
For more details on this class of bivariate PH distribution and an
extension to the IPH framework, we refer to \cite{albrecher2020fitting}.

The tractable class of mIPH distributions (with a particular instance, the mPH class)  was
introduced in \cite{bladt2022mph} by considering $d$ time-inhomogeneous Markov
pure jump processes on a common state space with $p$ transient states
that start in the same state and then evolve independently until
respective absorptions. If $Y^{(j)}$, $j=1,\dots,d$, are the univariate IPH
distributed absorption times of these Markov jump processes, we say that
the vector $\mathbf{Y}=( Y^{(1)},\ldots, Y^{(d)} )$ has a multivariate
inhomogeneous phase-type distribution of the mIPH type, and we write 
$$
\mathbf{Y}\sim \mbox{mIPH}(\boldsymbol{\alpha},\widetilde{S},\widetilde{L}), \quad \mbox{where}\quad \widetilde{S}=\{\mathbf{S}_1,\dots,\mathbf{S}_d\}\quad \text{and} \quad\widetilde{L}=\{\lambda_1,\dots,\lambda_d\}.
$$

This construction yields explicit expressions for important
functionals. For instance, let
$\mathbf{Y}\sim \mbox{mIPH}(\boldsymbol{\alpha},\widetilde{S},\widetilde{L})$,
then we have
$$
  f_\mathbf{Y}(\mathbf{y})=\sum_{k=1}^p \alpha_k \prod_{j=1}^d\mathbf{e}_k^{\top}\exp\left(\mathbf{S}_j g_j^{-1}(y^{(j)})\right)\mathbf{s}_j\lambda_j(y^{(j)}), \quad \mathbf{y}\in\mathbb{R}^d_+\,.
$$

Note that this allows for combining different types of
IPH marginals, given by $\lambda_i(\cdot)$ and $g_i(\cdot)$, which
enables different tail behaviours for the marginals.

\subsection{Regression}

For some of the multivariate classes introduced earlier, regression in
the vector of initial probabilities can be applied by utilizing the same
principle as in the univariate mixture-of-experts (MoE) approach. This
allows us to capture the relationships between the multivariate responses
and covariates, while taking advantage of the flexibility and
expressiveness of the multivariate matrix distributions. Specifically,
the mIPH-MoE model was introduced in \cite{albrecher2022joint} and used to
estimate the joint remaining lifetimes of spouses. Additionally, the
mDPH-MoE and fMDPH-MoE specifications were derived in \cite{bladt2023robust}
and illustrated for modelling insurance frequencies.

\section{Note on estimation algorithms and computational remarks} \label{sec:fit}

Most of the distribution classes presented above can be estimated using
various forms of expectation-maximisation (EM) algorithms. The original
EM algorithm for PH distribution was introduced in \cite{asmussen1996fitting}
and later extended for censored data in \cite{olsson1996estimation}. The
corresponding algorithm for DPH distributions can be found in
\cite{bladt2017matrix}. More recently, \cite{albrecher2020fitting} derived an EM
algorithm for IPH distributions and illustrated how the implementation
of the algorithms for MPH* and fMPH distributions can be carried out
smoothly. Regarding the regression models, the estimation method for the
PI model is presented in \cite{albrecher2022mortality}, while \cite{bladtyslas2022}
provides the algorithm for the estimation of the PH-MoE specification.
This method was then adapted in \cite{albrecher2022joint} for the mIPH class
and in \cite{bladt2023robust} for the DPH framework, which includes extensions
to the multivariate classes mDPH and fMDPH.

All estimation procedures implemented in the \pkg{matrixdist} package
related to PH distributions require an effective method for computing
matrix-exponential operations, as the various E-steps involved
hinge on these types of quantities. Several theoretical methods exist for these
purposes (see for instance \cite{moler1978nineteen}), but an efficient numerical implementation is
crucial due to the significant computational burden of the algorithms.
The \pkg{matrixdist} package offers three primary implementations of
matrix exponentials: the first, described in \cite{asmussen1996fitting},
involves converting the problem into a system of ordinary differential equations. The second method
utilises the so called uniformisation, with the exact description available in
\cite{albrecher2020fitting}. The third and final method is the Pad\'e
approximation, which can be found in \cite{moler1978nineteen}. Each method
presents its own advantages and disadvantages, and the choice will
depend on the specific problem/distribution at hand.

\section{Applications}\label{sec:ill}

 In this section, we illustrate the use of the  \texttt{matrixdist}  package in three actuarial applications of significance. The first example (Section~\ref{sec:autobi}) addresses the univariate modelling of loss data exhibiting heavy tails. The second example (Section~\ref{sec:mph}) deals with the joint modelling of absolute log returns through MPH* distributions. The final example (Section~\ref{sec:wisconsin}) focuses on a multiline insurance claim dataset, addressing both the marginal and joint modeling of claim frequency and severity across two distinct lines of insurance coverage.

\subsection{Automobile bodily injury claims}\label{sec:autobi}


We consider the \texttt{AutoBi} dataset from the \pkg{insuranceData} package,
which contains information about automobile bodily injury claims. The
dataset comprises eight variables: one for the economic loss (which we
aim to describe), six covariates, and one variable to identify each
claim. We start by providing general descriptive statistics for the
variable of interest (\texttt{LOSS}).

\begin{verbatim}
# Load the data
data("AutoBi", package = "insuranceData")
claim <- AutoBi
# Economic losses resulting from an automobile accident
loss <- claim$LOSS
# Summary statistics of losses
summary(loss)
\end{verbatim}

\begin{verbatim}
#>     Min.  1st Qu.   Median     Mean  3rd Qu.     Max. 
#>    0.005    0.640    2.331    5.954    3.995 1067.697
\end{verbatim}

\begin{verbatim}
# 10 largest values of losses
tail(sort(loss), n = 10)
\end{verbatim}

\begin{verbatim}
#>  [1]   82.000   96.007  114.604  150.000  162.047  188.720  193.000
#>  [8]  222.405  273.604  1067.697
\end{verbatim}

We observe that the losses range from 0.005 to 1067.697, with 75\% of the
data smaller than 3.995. The presence of large values, compared to the
body of the distribution, serves as an initial indication of a heavy
tail. To further confirm this observation, we create a Hill plot using
the \texttt{hill()} function from the \pkg{evir} package.

\begin{verbatim}
library(evir)
hill(loss)
\end{verbatim}

\begin{figure}[!htbp]
\centering 
\includegraphics[width=0.6\linewidth]{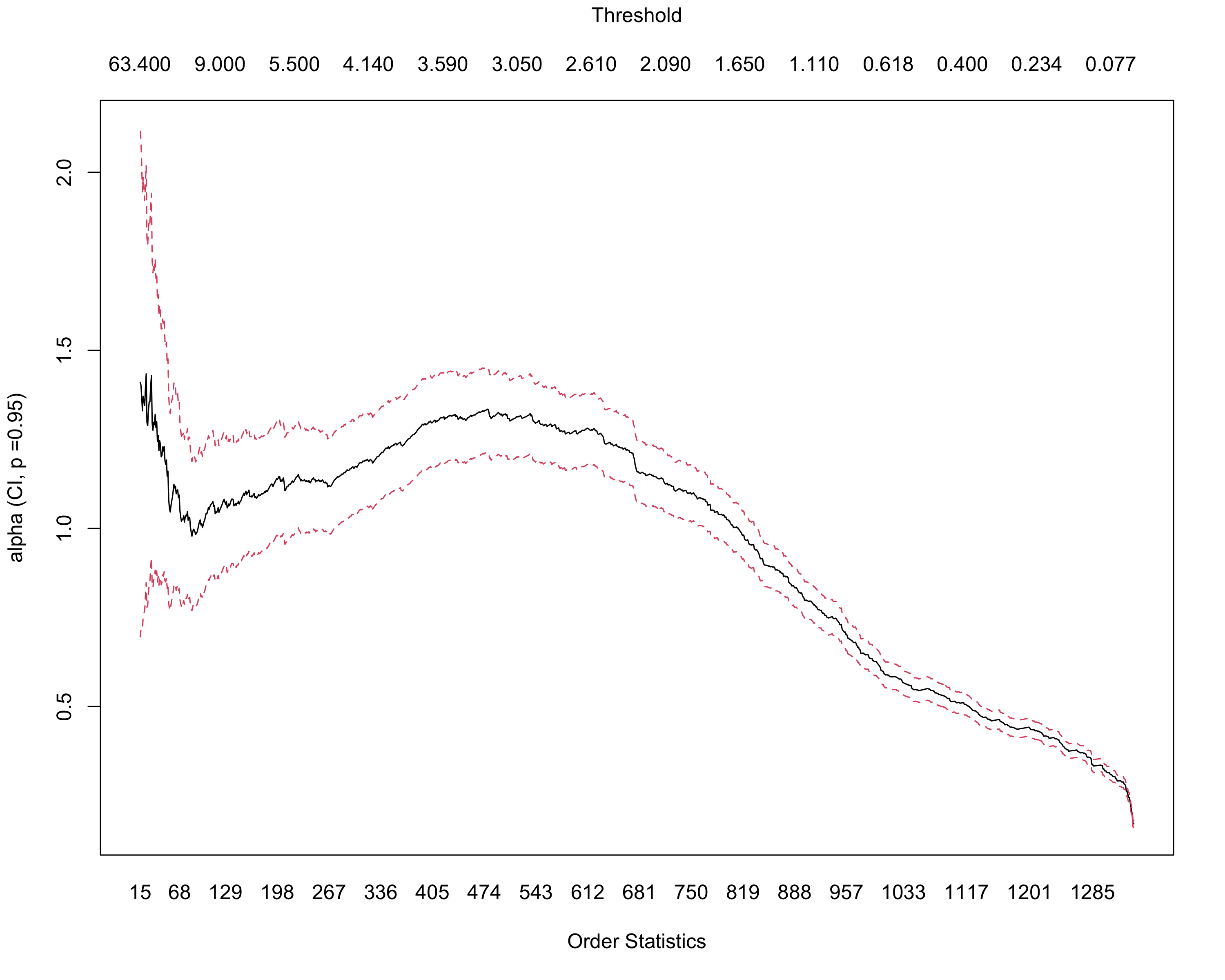} 
\caption{Hill plot for losses in the AutoBi dataset.}
\label{fig:hillplot}
\end{figure}

The plot exhibits a flat behaviour for the largest order statistics, suggesting a Pareto-type tail (or regularly varying). Therefore, we opt for an IPH model to more accurately capture the tail of the losses. More specifically, we use a matrix-Pareto distribution. The fitting process starts by creating an initial IPH distribution, which provides the starting parameters for the EM algorithm. This can be accomplished by employing the \texttt{iph()} method in \pkg{matrixdist}. For our analysis, we create an IPH object with random parameters of a general structure, a dimension 6, and a Pareto inhomogeneity function.

\begin{verbatim}
# Initial IPH object
set.seed(22)
x <- iph(gfun = "pareto", structure = "general", dimension = 6)
\end{verbatim}

With an initial IPH distribution in place, we can employ the \texttt{fit()} method to fit the data. It is important to note that \texttt{fit()} requires specifying the number of steps for the EM algorithm. In this instance, we use 1500 steps, as the changes in the loglikelihood become negligible with this number (the same criteria is employed to determine the number of iterations in the subsequent illustration). The method outputs an IPH object containing the fitted parameters. Further steps could be made on the output object if required.

\begin{verbatim}
# Fitting procedure
z <- fit(x, y = loss, stepsEM = 1500) 
\end{verbatim}

\begin{figure}[!htbp]
\centering 
\includegraphics[width=0.6\linewidth]{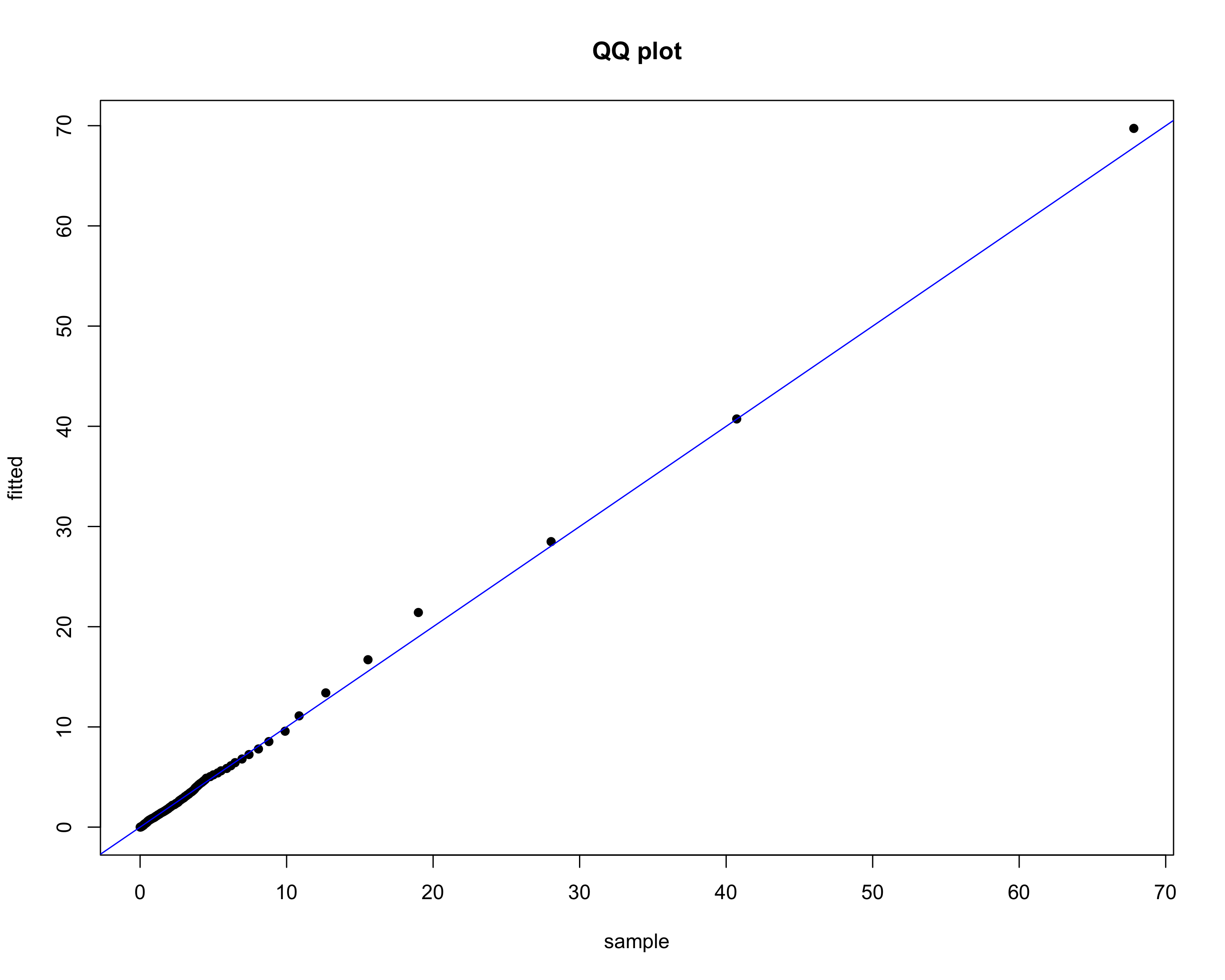}
\caption{QQ plot of fitted distribution.}
\label{fig:uniPHplot}
\end{figure}

\begin{figure}[!htbp]
\centering \includegraphics[width=0.6\linewidth]{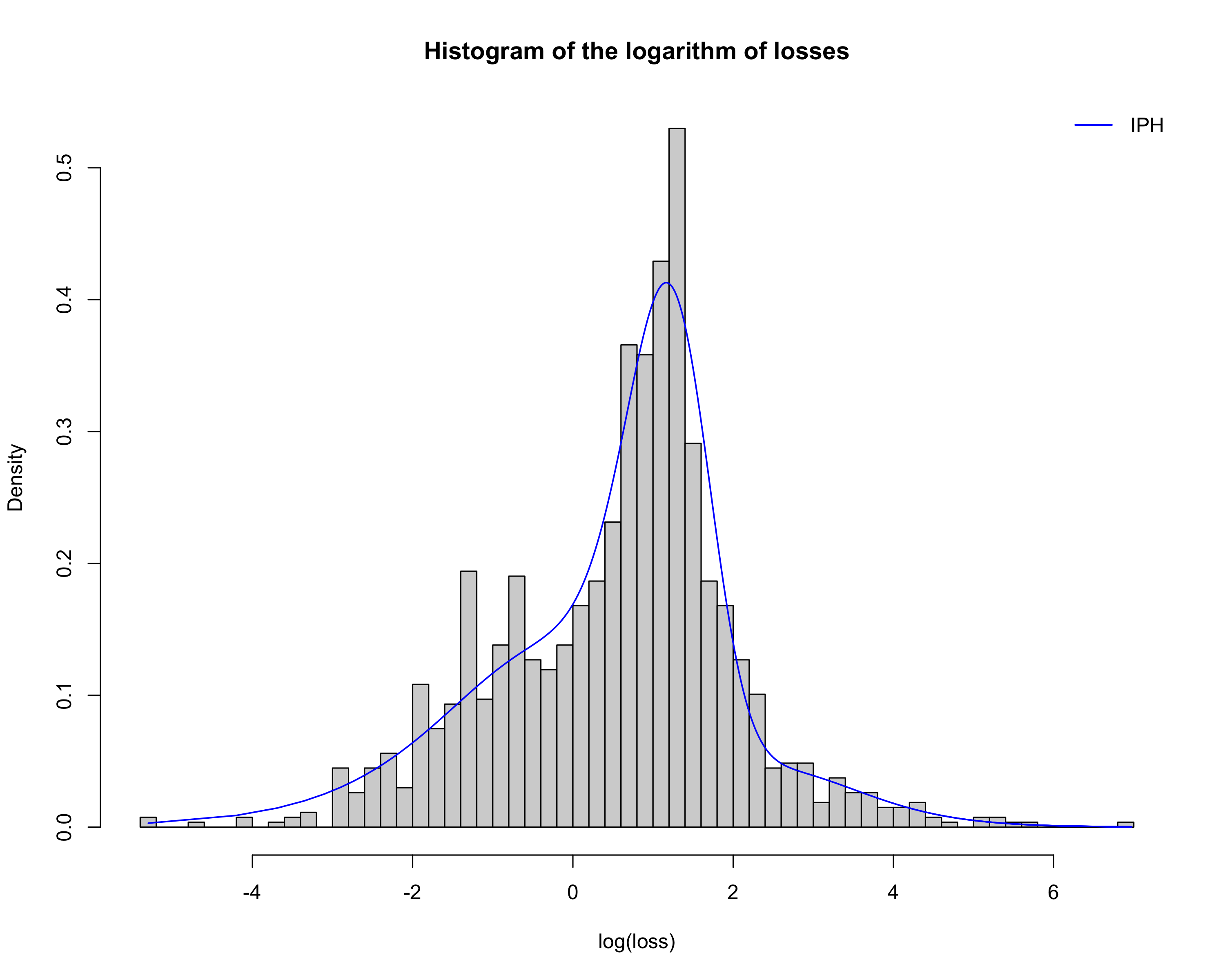} 
\caption{Histogram of the logarithm of losses vs fitted density (blue line).}
\label{fig:IPHhist}
\end{figure}

Figures \ref{fig:uniPHplot} and \ref{fig:IPHhist} display a QQ plot for the fitted distribution and a histogram of the (log) data against the fitted density, respectively. These plots indicate that the fitted distribution serves as a good approximation of the data. In particular, the sample and fitted quantiles align closely with the identity line, especially within the distribution's body. Although there is a slight deviation from the identity line for quantiles in the tail, they remain relatively close to the blue line. The histogram further supports the notion that the fitted IPH distribution is a satisfactory approximation.

\subsubsection*{PH regression}\label{ph-regression}

We now shift our focus to the regression framework. In particular, we employ the PI specification to capture the influence of covariate information on economic losses. The initial step involves removing any incomplete covariate data. Next, we need to split the covariates and the response variable into two different objects. In our case, we investigate the effect of the variables \texttt{ATTORNEY}, \texttt{CLMSEX} and \texttt{CLMAGE}. The variable \texttt{ATTORNEY} indicates whether the claimant was represented by an attorney or not, \texttt{CLMSEX} is the claimant's sex, and \texttt{CLMAGE} corresponds to their age.

\begin{verbatim}
# Covariate information, excluding missing entries
cov <- na.omit(claim)
# Separation of the information
X <- cov[, c("ATTORNEY", "CLMSEX", "CLMAGE")]
loss <- cov[, "LOSS"]
\end{verbatim}

Now, we can employ the \texttt{reg()} method to perform the PI regression. This function works similarly to the \texttt{fit()} method, with the main difference being that estimation of the regression parameters is included in each iteration of the EM algorithm (see \cite{albrecher2022mortality} for details). Along with the data, the \texttt{reg}() method requires an initial IPH distribution. In this case, we employ the previously fitted matrix-Pareto distribution without covariates.

\begin{verbatim}
# PI regression
z.reg <- reg(x = z, y = loss, X = X, stepsEM = 250) 
\end{verbatim}

The resulting object (\texttt{z.reg}) contains all matrix parameters along with the regression coefficients. This information can be easily accessed via the \texttt{coef()} function as follows.

\begin{verbatim}
# Fitted parameters
z.reg.par <- coef(z.reg)
z.reg.par
\end{verbatim}

\begin{verbatim}
#> $alpha
#> [1] 2.356094e-04 3.257751e-06 5.920828e-02 3.755407e-04 4.561202e-01
#> [6] 4.840571e-01
#> 
#> $S
#>               [,1]          [,2]        [,3]          [,4]          [,5]
#> [1,] -3.412707e+01  7.592078e-06  1.42774265  1.533919e-01  6.975181e-06
#> [2,]  2.145191e+01 -3.473845e+01  0.38024154  1.288647e+01  3.162116e-05
#> [3,]  3.090083e-01  6.287658e-01 -1.83234510  4.643455e-01  3.670351e-01
#> [4,]  5.666353e-01  1.166288e-04  7.77223071 -2.987094e+01  2.335434e-04
#> [5,]  1.432727e-04  3.578715e+01  0.49172745  2.247409e-04 -3.628114e+01
#> [6,]  4.561651e-06  1.243609e-03  0.08311972  3.771681e-05  4.707313e-02
#>               [,6]
#> [1,]  3.037507e+01
#> [2,]  2.799554e-05
#> [3,]  3.352413e-02
#> [4,]  2.066459e+01
#> [5,]  1.118600e-03
#> [6,] -6.641175e+00
#> 
#> $beta
#> [1] 33.10073
#> 
#> $B
#>                                     
#>  1.15139949 -0.10888748 -0.01368395 
#> 
#> $C
#> numeric(0)
\end{verbatim}

The above information can be used, for example, to create customised loss distributions for individual claimants based on their distinctive characteristics.

Take, for instance, a new claimant: a 60-year-old male, unrepresented by an attorney. The following code leverages our model to estimate his loss distribution.

\begin{verbatim}
# Covariate information of the new claimant
claimant <- data.frame(ATTORNEY = 2, CLMSEX = 1, CLMAGE = 60)
# Multiplicative effect in the sub-intensity matrix
prop <- exp(sum(z.reg.par$B * claimant))
# IPH distribution for claimant's loss
claimant.iph <- iph(
  alpha = z.reg.par$alpha, S = z.reg.par$S * prop,
  gfun = "pareto", gfun_pars = z.reg.par$beta
)
\end{verbatim}

If we wish to visualize this claimant's survival function, we can plot it as follows.

\begin{verbatim}
# Evaluation points for the survival function
q <- seq(0, 15, 0.01)
# Plot of the survival function
plot(q, cdf(claimant.iph, q, lower.tail = F),
  type = "l", col = "blue", lwd = 1.3,
  ylab = "1-F(y)", xlab = "y", main = "Survival function"
)
\end{verbatim}

\begin{figure}[!htbp]
\centering \includegraphics[width=0.6\linewidth]{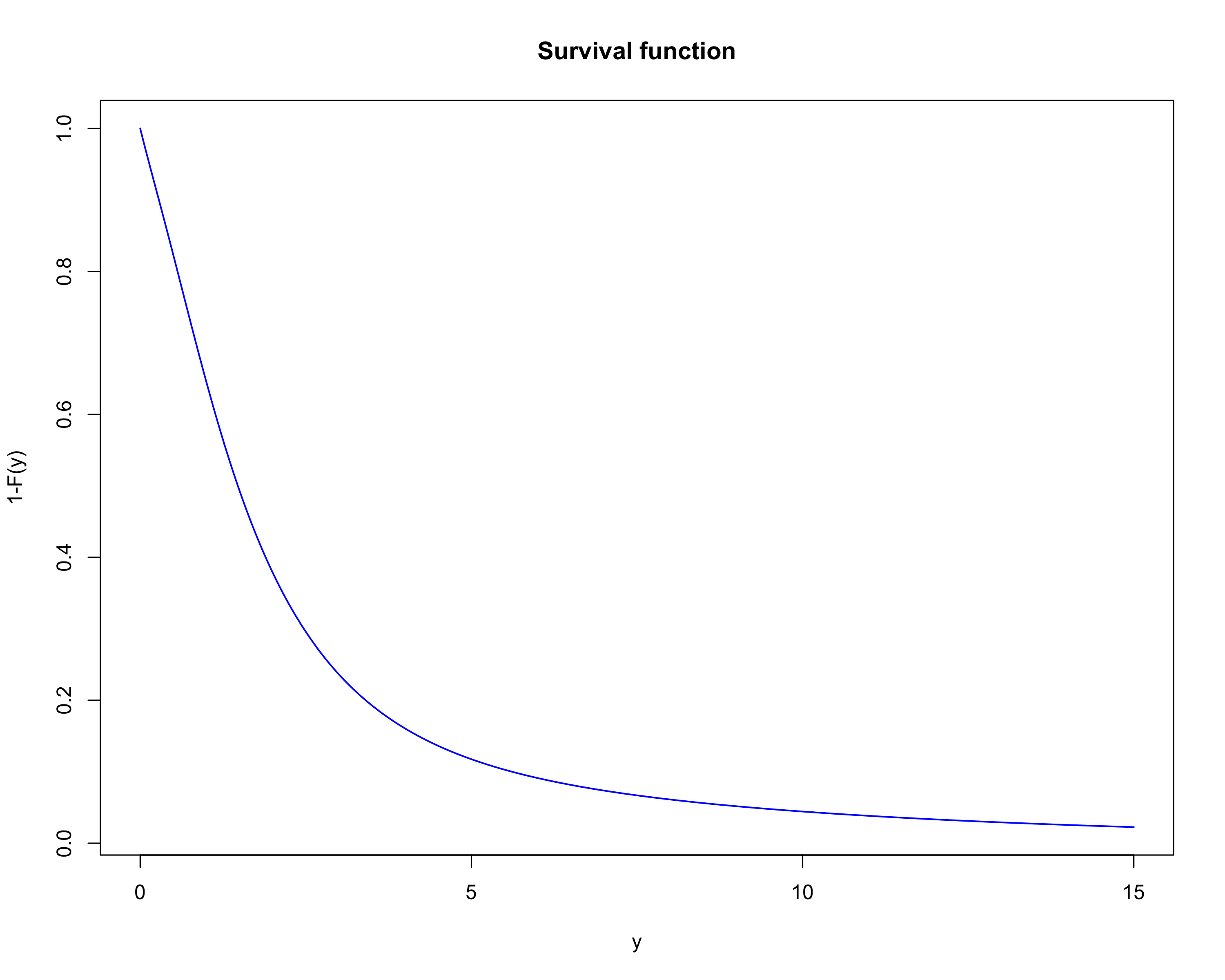} 
\caption{Survival function of losses for the new claimant.}\label{fig:PHregSurv}
\end{figure}

We refer readers to Section~\ref{sec:sum} and the documentation of \pkg{matrixdist} for further available methods for PH and IPH distributions.

\subsection{Multivariate modelling of log returns}\label{sec:mph}


To illustrate the capabilities of the implementations for MPH* distributions in \pkg{matrixdist}, we use the \texttt{rdj} dataset of the \pkg{copula} package. More specifically, our focus is to use an MPH* for describing the joint behaviour of the absolute log returns of the three stock prices in this dataset: Intel, Microsoft, and General Electric.

Firstly, we load the data, take the absolute value, and eliminate any data points with atoms at zero.

\begin{verbatim}
set.seed(24)
# Load the data
data("rdj", package = "copula")
# Absolute log return of stock prices
y <- as.matrix(abs(rdj[, -1]))
# Remove data points with atoms at zero
y <- y[y[, 1] != 0 & y[, 2] != 0 & y[, 3] != 0, ]
\end{verbatim}

As in the univariate case, the fitting process requires the definition of an initial \texttt{MPHstar} object that is subsequently fed into the \texttt{fit()} function. In the present case, we use an \texttt{MPHstar} object with randomly generated parameters, a general form of the matrix parameters, and dimension \(10\).

\begin{verbatim}
# Initial MPH* distribution
x <- MPHstar(structure = "general", dimension = 10, variables = 3)
# Fitting to data
x.fit <- fit(x = x, y = y, stepsEM = 200, uni_epsilon = 1e-6, r = 0.5)
\end{verbatim}

In particular, the reward matrix of the fitted distribution can be accessed as follows:

\begin{verbatim}
# Estimated reward matrix
x.fit@pars$R
\end{verbatim}

\begin{verbatim}
#>              [,1]       [,2]       [,3]
#>  [1,] 0.624656883 0.03162784 0.34371527
#>  [2,] 0.448807782 0.08398967 0.46720255
#>  [3,] 0.545561519 0.42181735 0.03262113
#>  [4,] 0.783031036 0.08518594 0.13178302
#>  [5,] 0.973586315 0.01054082 0.01587287
#>  [6,] 0.318181076 0.21878855 0.46303038
#>  [7,] 0.015947222 0.78138202 0.20267076
#>  [8,] 0.498338747 0.28584901 0.21581224
#>  [9,] 0.001025196 0.99897480 0.00000000
#> [10,] 0.471608399 0.48276616 0.04562544
\end{verbatim}

Finally, we assess the quality of the fit by creating QQ plots of the fitted margins.

\begin{verbatim}
# Names for the QQ plot
name <- colnames(y)
# Quantiles' levels
qq <- seq(0.01, 0.99, length.out = 100)
par(mfrow = c(1, 3))
for (j in 1:3) {
  # j-th marginal fitted distribution
  x <- marginal(x.fit, j)
  # Marginal observations
  ym <- y[, j]
  # quantiles
  y.q <- quantile(ym, probs = qq)
  x.q <- quan(x, qq)
  # QQ plot
  plot(y.q, x.q, main = name[j], xlab = "Sample", 
    ylab = "Fitted", pch = 16, asp = 1
  )
  abline(0, 1, col = "blue")
}
\end{verbatim}

\begin{figure}[!htbp]
{\centering \includegraphics[width=0.85\linewidth]{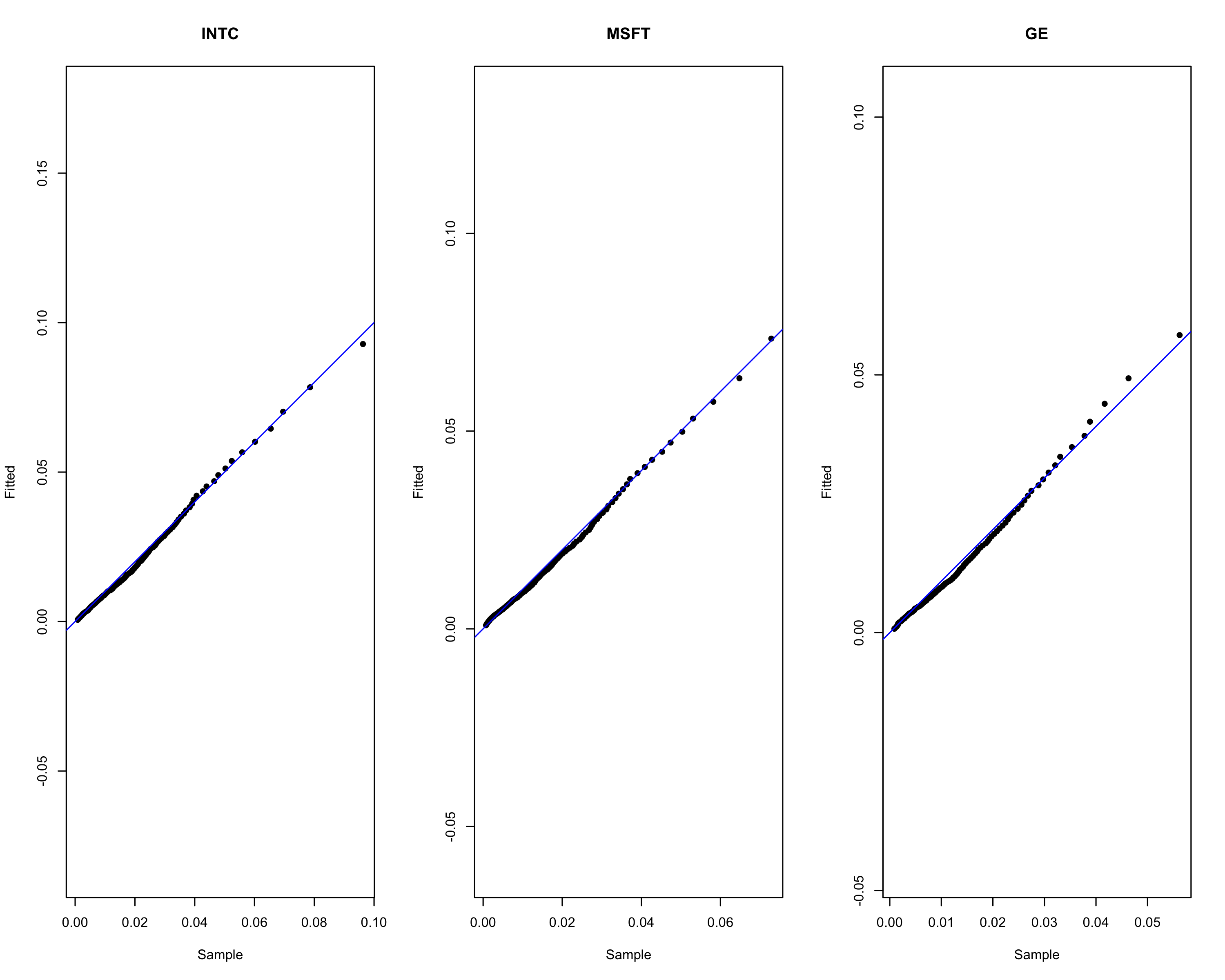} 
}
\caption{QQ plot of fitted distirbutions for each marginal.}\label{fig:qqMPHstar}
\end{figure}

Upon inspecting the QQ plots (Figure~\ref{fig:qqMPHstar}), we can observe that the fitted marginals provide a satisfactory approximation of the data.

\subsection{Wisconsin Property Insurance dataset}\label{sec:wisconsin}

We now consider the Wisconsin Local Government Property Insurance Fund (LGPIF) dataset (see \url{https://sites.google.com/a/wisc.edu/local-government-property-insurance-fund}). This fund was established to provide insurance coverage to government properties not owned by the State of Wisconsin. Governmental entities under this fund include counties, cities, towns, schools, and other miscellaneous entities, and properties covered encompass buildings, vehicles, and equipment. The dataset is divided into a training set covering the years 2006-2010 and a testing set encompassing the year 2011. It includes data on claim frequencies and severities across six different coverage groups: building and content (BC), contractor's equipment (IM), comprehensive new (PN), comprehensive old (PO), collision new (CN), and collision old (CO) coverage.

 This dataset, which is publicly available at \url{https://sites.google.com/a/wisc.edu/jed-frees/home}, was thoroughly studied in  \cite{frees2016multivariate}, with the corresponding code used for the analysis also accessible at the same URL. Here, we focus on the univariate and multivariate modelling of frequency and severity for two coverages -- BC and CO -- using matrix models and compare our results to those obtained in the aforementioned study.

\subsubsection*{Severity modelling}
We start our analysis with the univariate modelling of the average claim size of BC, which we scale with a factor of $10^{-4}$ to speed up the convergence of the fitting algorithms.
\begin{verbatim}
sevinBC$yAvgBC <- sevinBC$yAvgBC * 1e-4	
\end{verbatim}

\begin{figure}[!htbp]
\centering \includegraphics[width=0.6\linewidth]{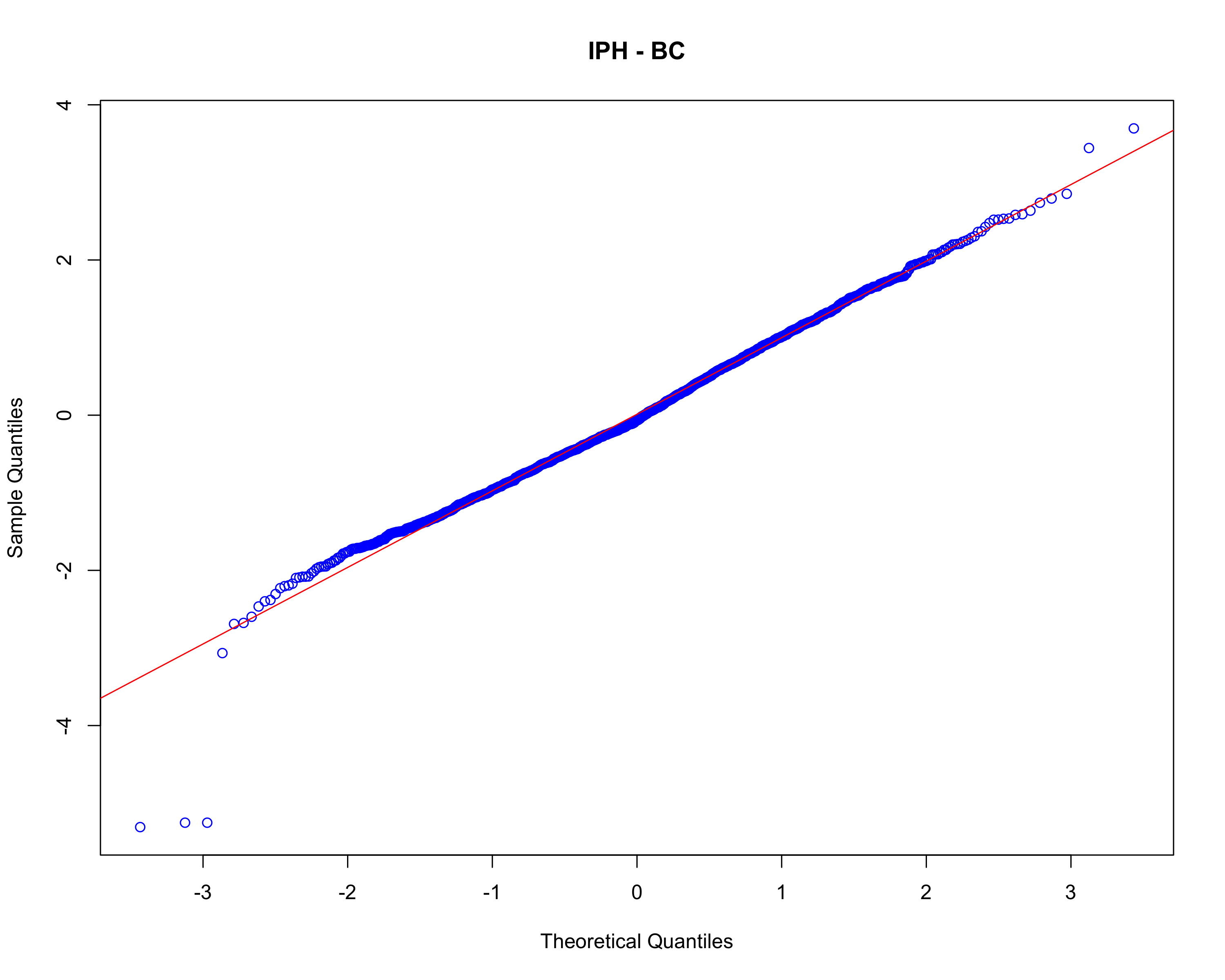} 
\caption{QQ Plot for residuals of IPH for BC.}\label{fig:cs_bc}
\end{figure}

Regarding explanatory variables, we choose the following as in  \cite{frees2016multivariate}: coverage amount (in log-scale), an indicator for no claims in the previous year, entity type (City, County, Misc, School, Town), and deductible level (in log-scale). We then fit a PH-MoE model of dimension 5 with general Coxian structure of the sub-intensity matrix. For this, we first specify the regression formula and an initial IPH model. 
\begin{verbatim}
# Regression formula
formula <- yAvgBC ~ CoverageBC + lnDeductBC + NoClaimCreditBC + TypeCity
+TypeCounty + TypeMisc + TypeSchool + TypeTown
# Initial IPH object
set.seed(5054)
x <- iph(ph(structure = "gcoxian", dimension = 5),
  gfun = "lognormal", gfun_pars = 1
)
\end{verbatim}
These two objects are then passed into the \texttt{MoE()} function to obtain the fitted PH-MoE model.
\begin{verbatim}
# MoE regression
phMoE_BC <- MoE(x = x, formula = formula, data = sevinBC, stepsEM = 1000)	
\end{verbatim}
Note that we have selected a matrix-lognormal model, as it provided the largest value of the loglikelihood ($-1,836.67$) compared to other IPH specifications of the same dimension. In fact, it outperforms the Generalised Beta type 2 (GB2) model in \cite{frees2016multivariate}, which has corresponding loglikelihood of $-1,973.30$. The goodness-of-fit is then checked visually using a QQ plot of the quantiles of normal Cox-Snell residuals, which are compared with normal quantiles (see Figure~\ref{fig:cs_bc}).
\begin{verbatim}
# Cox-Snell residuals
cs <- rep(0, length(sevinBC$yAvgBC))
for (k in 1:length(sevinBC$yAvgBC)) {
  cs[k] <- cdf(iph(ph(phMoE_BC$alpha[k, ], phMoE_BC$S),
    gfun = "lognormal", gfun_pars = phMoE_BC$inhom$pars
  ), sevinBC$yAvgBC[k])
}
# QQ-plot of residuals
qqnorm(qnorm(cs), col = "blue", main = "IPH - BC")
qqline(qnorm(cs), col = "red")	
\end{verbatim}

Moreover, to assess the predictive performance of the fitted model, we use the testing set and compared the mean predictions of the PH-MoE and GB2 models with the observed average severity via the root-mean-square error (RMSE) measure. 
\begin{verbatim}
# Testing set
sevoutBC <- dataout[BCpout, ]
nBC <- length(BCpout)
# Prediction for vector of initial probabilities
alpha_vecs <- stats::predict(phMoE_BC$mm, type = "probs", newdata = sevoutBC)
# Mean values
mean_val <- c()
for (j in 1:nBC) {
  iph_temp <- iph(ph(alpha_vecs[j, ], phMoE_BC$S),
    gfun = "lognormal", gfun_pars = phMoE_BC$inhom$pars
  )
  mean_val[j] <- integrate(function(x) x * dens(iph_temp, x), 0, Inf)$value
}
# RMSE
sqrt(sum((mean_val - sevoutBC$yAvgBC * 1e-04)^2) / nBC)
\end{verbatim}

We obtain an RMSE of 7.8305 for the PH-MoE model and 29.8380 for the GB2 one, showcasing the superior predictive performance of the matrix model.

Subsequently, we perform a similar analysis for the CO coverage. In this case we only consider coverage amount (in log-scale) and the indicator for no claims in the previous year as covariates to align the results with the study in \cite{frees2016multivariate}. Moreover, we have also selected a matrix-lognormal model of dimension 4 and general Coxian structure of the sub-intensity matrix. 
\begin{verbatim}
# Scaling of average claim sizes
sevinCO$yAvgCO <- sevinCO$yAvgCO * 1e-04
# Regression formula
formula <- yAvgCO ~ CoverageCO + NoClaimCreditCO
# Initial IPH object
set.seed(8919)
x <- iph(ph(structure = "gcoxian", dimension = 4),
  gfun = "lognormal", gfun_pars = 1
)
# MoE regression
phMoE_CO <- MoE(x = x, formula = formula, data = sevinCO, stepsEM = 1000)
\end{verbatim}

\begin{figure}[!htbp]
\centering \includegraphics[width=0.6\linewidth]{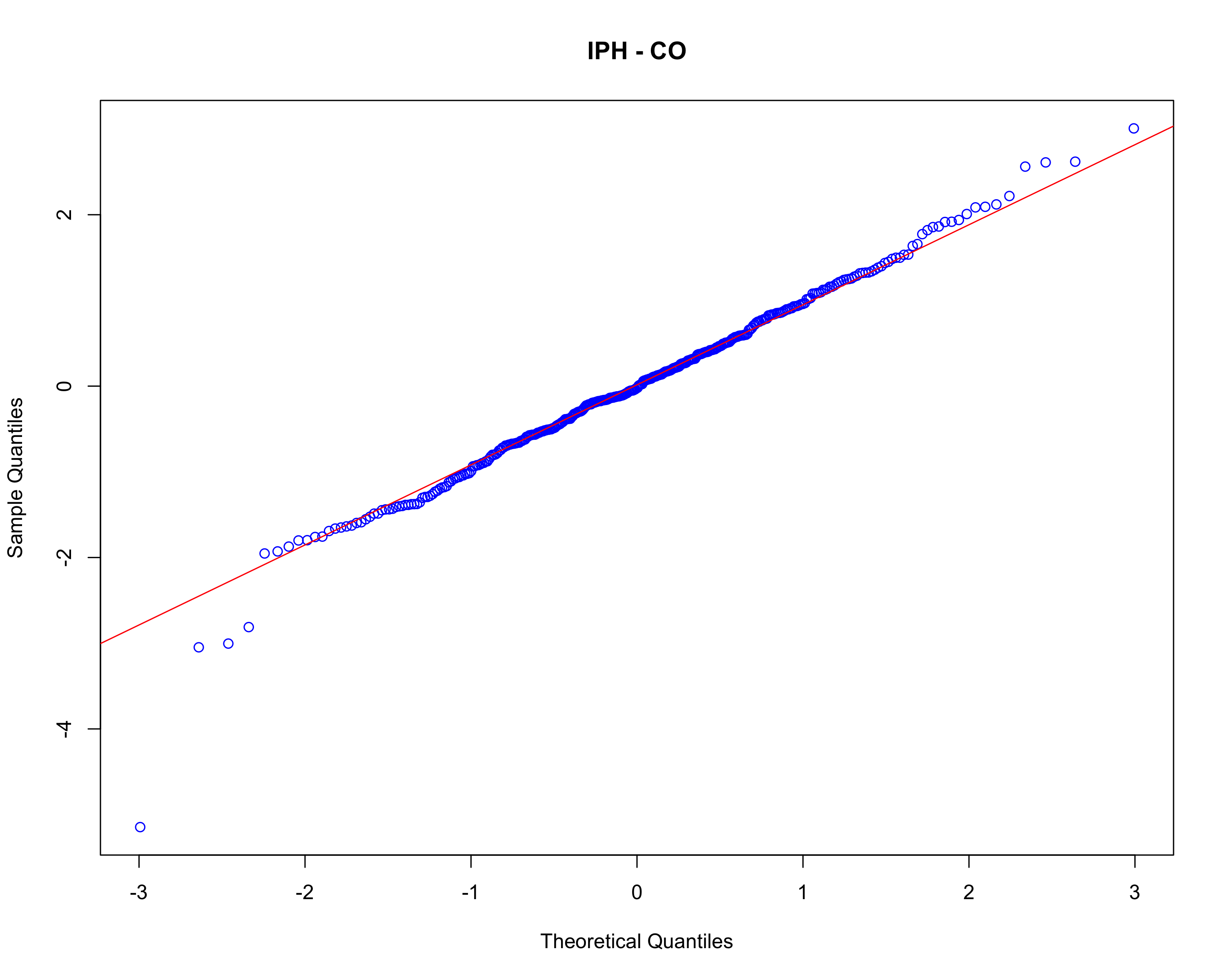} 
\caption{QQ Plot for residuals of IPH for CO.}\label{fig:cs_co}
\end{figure}

Once again, we have that the matrix model outperforms the GB2 model in terms of loglikelihood with -178.78 for the former and -189.38 for the latter. The QQ plot of the residuals of the PH-MoE model are shown in Figure~\ref{fig:cs_co}.
\begin{verbatim}
# Cox-Snell residuals
cs <- rep(0, length(sevinCO$yAvgCO))
for (k in 1:length(sevinCO$yAvgCO)) {
  cs[k] <- cdf(iph(ph(phMoE_CO$alpha[k, ], phMoE_CO$S),
    gfun = inh_name, gfun_pars = phMoE_CO$inhom$pars
  ), sevinCO$yAvgCO[k])
}
# QQ-plot of residuals
qqnorm(qnorm(cs), col = "blue", main = "IPH - CO")
qqline(qnorm(cs), col = "red")	
\end{verbatim}

Regarding the predictive performance, the PH-MoE specification, with an RMSE  of 1.1552, outperforms the GB2 model with an RMSE  of 1.2277.  
\begin{verbatim}
# Testing set
sevoutCO <- dataout[COpout, ]
nCO <- length(COpout)
# Prediction for vector of initial probabilities
alpha_vecs <- stats::predict(phMoE_CO$mm, type = "probs", newdata = sevoutCO)
# Mean values
mean_val <- c()
for (j in 1:nCO) {
  iph_temp <- iph(ph(alpha_vecs[j, ], phMoE_CO$S),
    gfun = "lognormal", gfun_pars = phMoE_CO$inhom$pars
  )
  mean_val[j] <- integrate(function(x) x * dens(iph_temp, x), 0, Inf)$value
}
# RMSE
sqrt(sum((mean_val - sevoutCO$yAvgCO * 1e-04)^2) / nCO)
\end{verbatim}

We conclude the analysis of the severities by presenting a joint modelling of the BC and CO lines using an mIPH model with MoE regression. We start by considering the data with claims in both lines and giving the same scaling of $10^{-4}$ as before. 
\begin{verbatim}
biv_ind <- which(data$ClaimBC > 0 & data$ClaimCO > 0 &
  data$CoverageCO > 0 & data$CoverageBC > 0)
data_biv <- data[biv_ind, ]

data_biv$yAvgCO <- data_biv$yAvgCO * 1e-04
data_biv$yAvgBC <- data_biv$yAvgBC * 1e-04
data_biv$CoverageBC <- log(data_biv$CoverageBC)
data_biv$CoverageCO <- log(data_biv$CoverageCO)

responses <- cbind(data_biv$yAvgBC, data_biv$yAvgCO)
\end{verbatim}
We then specify the regression formula and an initial mIPH model. In this case, we have chosen matrix-lognormal margins of dimension 5 and general Coxian structure of their sub-intensity matrices, primarily based on the previous univariate analysis.
\begin{verbatim}
# Regression formula
formula_biv <- responses ~ CoverageCO + NoClaimCreditCO + CoverageBC
+lnDeductBC + NoClaimCreditBC + TypeCity + TypeCounty + TypeMisc + TypeSchool
+TypeTown
# Initial mIPH object
set.seed(741)
x1 <- mph(structure = c("gcoxian", "gcoxian"), dimension = 5)
x2 <- miph(x1, gfun = c("lognormal", "lognormal"), gfun_pars = list(c(1), c(1)))
\end{verbatim}

Finally, we fit the regression model using the \texttt{MoE()} function. 
\begin{verbatim}
# MoE regression
bivMoE <- MoE(
  x = x2,
  formula = formula_biv,
  y = responses,
  data = data_biv,
  stepsEM = 500
)
\end{verbatim}
We note that the fitted model gives a value of the loglikelihood of -332.79, which outperforms the joint model in \cite{frees2016multivariate}, consisting of GB2 margins bounded together with a Gaussian copula, with a corresponding loglikelihood of -341.33.

\subsubsection*{Frequency modelling}
As in the previous section, we begin our study with the univariate modelling of the frequency of BC. Since the support of DPH distributions is $\mathbb{N}$ and the dataset includes observations equal to zero, we apply a translation to the data to start at one.
\begin{verbatim}
# Translation
dataBC <- freqinBC
dataBC$FreqBC <- dataBC$FreqBC + 1
\end{verbatim}

We now specify the regression formula, which includes the same covariates as the corresponding analysis in \cite{frees2016multivariate}, and an initial DPH object of dimension 7 with a general Coxian-like structure of the matrix of transition probabilities. 
\begin{verbatim}
# Regression formula
formula <- FreqBC~ CoverageBC + lnDeductBC + NoClaimCreditBC + TypeCity
+TypeCounty + TypeMisc + TypeSchool + TypeTown
# Initial DPH object
set.seed(1212)
x <- dph(structure = "gcoxian", dimension = 7)
\end{verbatim}

Regression under the MoE framework can now be performed using the above two objects in the \texttt{MoE()} function. 
\begin{verbatim}
# MoE regression
dphMoE_BC <- MoE(x = x, formula = formula, data = dataBC, stepsEM = 500)
\end{verbatim}
The fitted model leads to a loglikelihood of -4,777.64, which is considerably higher than the corresponding values of the two models with the largest loglikelihoods in \cite{frees2016multivariate}; negative binomial (NB) with a value of -5,102.62 and zero-one-inflated negative binomial (zerooneNB) with -4,958.29. To further evaluate the goodness-of-fit, we calculate the chi-square statistic, which yields a value of 22.20 for the DPH-MoE model. This outperforms the NB and zerooneNB specifications with corresponding values of 88.09 and  34.51, respectively.
\begin{verbatim}
# Chi-square statistic
numrow <- max(freqBC) + 1
eDPH <- 0
for (j in 1:nrow(dataBC)) {
  eDPH <- eDPH + dens(dph(dphMoE_BC$alpha[j, ], dphMoE_BC$S), 1:numrow)
}
sum((c(eDPH[1:19], sum(eDPH[20:numrow])) - c(emp[1:19], sum(emp[20:numrow])))^2
  / c(eDPH[1:19], sum(eDPH[20:numrow])))
\end{verbatim}

Regarding the predictive performance, we use the testing set and compare the mean predictions of the DPH-MoE, NB, and zerooneNB models with the observed frequencies using the RMSE. 
\begin{verbatim}
# Testing set
dataBCout <- doutBC
dataBCout$FreqBC <- dataBCout$FreqBC + 1
nBC <- nrow(doutBC)
# Prediction for vector of initial probabilities
alpha_vecs <- stats::predict(dphMoE_BC$mm, type = "probs", newdata = dataBCout)
# Mean values
mean_val <- c()
for (j in 1:nBC) {
  mean_val[j] <- mean(dph(alpha_vecs[j, ], dphMoE_BC$S)) - 1
}
# RMSE
sqrt(sum((mean_val - doutBC$FreqBC)^2) / nBC)
\end{verbatim}
We have that the DPH-MoE specification outperforms the other two models with a RMSE of 6.1806. In contrast, the NB model yields a RMSE of 6.8581, while the zerooneNB model shows a value of 6.9722.

We continue the study with the univariate modelling of the frequency for the CO line. In this case, we have selected a DPH-MoE model of dimension 6 with the same sparse structure of the transition matrix. 
\begin{verbatim}
# Translation
dataCO <- freqinCO
dataCO$FreqCO <- dataCO$FreqCO + 1
# Regression formula
formula <- FreqCO~ CoverageCO + NoClaimCreditCO + TypeCity + TypeCounty
+ TypeMisc + TypeSchool + TypeTown
# Initial DPH object
set.seed(8)
x <- dph(structure = "gcoxian", dimension = 6)
# MoE regression
dphMoE_CO <- MoE(x = x, formula = formula, data = dataCO, stepsEM = 1000)
\end{verbatim}
We have that the fitted DPH-MoE yields a value of the loglikelihood of -1,022.96, which is larger than the corresponding values of the NB and zerooneNB models with values of -1071.140 and  -1067.823, respectively. However, note that the value of the chi-square statistics is 10.49, which is marginally bigger than the values of 10.39 and 10.37 coming from the NB and zerooneNB specifications. 
\begin{verbatim}
# Chi-square statistic
numrow <- max(freqCO) + 1
eDPH <- 0
for (j in 1:nrow(dataCO)) {
  eDPH <- eDPH + dens(dph(dphMoE_CO$alpha[j, ], dphMoE_CO$S), 1:numrow)
}
sum((c(eDPH[1:10], sum(eDPH[11:numrow])) - c(emp[1:10], sum(emp[11:numrow])))^2
    / c(eDPH[1:10], sum(eDPH[11:numrow])))
\end{verbatim}

Moreover, we assess the predictive performance of the fitted DPH-MoE model using the testing set and the RMSE measure. We obtain a value of 0.5847 for this measure, which indicates superior performance compared to the NB and zerooneNB specifications, with corresponding values of 0.6615 and 0.6110.
\begin{verbatim}
dataCOout <- doutCO
dataCOout$FreqCO <- dataCOout$FreqCO + 1
nCO <- nrow(doutCO)
# Prediction for vector of initial probabilities
alpha_vecs <- stats::predict(dphMoE_CO$mm, type = "probs", newdata = dataCOout)
# Mean values
mean_val <- c()
for (j in 1:nCO) {
  mean_val[j] <- mean(dph(alpha_vecs[j, ], dphMoE_CO$S)) - 1
}
# RMSE
sqrt(sum((mean_val - doutCO$FreqCO)^2) / nCO)
\end{verbatim}

Finally, we focus on the joint modelling of the frequencies for the BC and CO lines via an mDPH model with MoE regression. We first need to filter the data to consider only the entires with coverage in both lines and apply a translation to one. 
\begin{verbatim}
biv_ind <- which(data$CoverageBC > 0 & data$CoverageCO > 0)
data_biv <- data[biv_ind, ]

data_biv$FreqCO <- data_biv$FreqCO + 1
data_biv$FreqBC <- data_biv$FreqBC + 1
data_biv$CoverageBC <- log(data_biv$CoverageBC)
data_biv$CoverageCO <- log(data_biv$CoverageCO)
responses <- cbind(data_biv$FreqBC, data_biv$FreqCO)
\end{verbatim}
We then specify the regression formula and an initial mDPH object. Note that we have selected dimension 7 with a general Coxian-like structure of the transition matrices in accordance with the previous univariate analyses. 
\begin{verbatim}
# Regression formula
formula <- responses ~ CoverageCO + NoClaimCreditCO + CoverageBC + lnDeductBC
+NoClaimCreditBC + TypeCity + TypeCounty + TypeMisc + TypeSchool + TypeTown
# Initial mDPH model
set.seed(12345)
x <- mdph(structure = c("gcoxian", "gcoxian"), dimension = 7)
\end{verbatim}
With these two objects at hand, we can now perform regression via the MoE framework using the \texttt{MoE()} function. 
\begin{verbatim}
# MoE regression
bivMoE <- MoE(
  x = x,
  formula = formula,
  y = responses,
  data = data_biv,
  stepsEM = 350
)
\end{verbatim}
The resulting fitted matrix model yields a loglikelihood of -2,979.958. This outperforms the multivariate model in \cite{frees2016multivariate}, consisting of a Gaussian copula with zerooneNB margin for BC and NB margin for CO, which has leads to a value of -3,321.726 of the loglikelihood.

\section{Summary of methods}\label{sec:sum}

We have explored the fundamental properties of some matrix distributions and
demonstrated the use of the \pkg{matrixdist} package for analyzing
real-life data through detailed examples. The package encompasses
methods for computing various functionals and estimating most of the
classes discussed in this paper. Table \ref{tab:classes} presents the
families of matrix distributions currently implemented in the package.
We refer readers to the package documentation for guidance on using
these construction functions. For example, consulting \texttt{?ph} will provide
instructions on creating PH distributions. Additionally, Table
\ref{tab:methods} offers a comprehensive list of the implemented
methods, along with a mapping to the applicable classes. For examples of
the use of these methods, the package help can be consulted. The general
structure for accessing the desired examples is
\texttt{?\textasciigrave{}method\_name,class\_name-method\textasciigrave{}}. For instance,
\texttt{?\textasciigrave{}fit,ph-method\textasciigrave{}} directs to the documentation of the \texttt{fit()}
method for the \texttt{ph} class. In future versions of the package, methods and classes are expected to gradually be included, including general heavy-tailed models as in \cite{albrecher2021cph, bladt2021heavyph}.

\begin{table}

\caption{\label{tab:classes}Classes of matrix distributions available in matrixdist}
\centering
\begin{tabular}[!htbp]{ll}
\toprule
Class & Constructor function\\
\midrule
DPH & dph()\\
PH & ph()\\
IPH & iph()\\
fMDPH & bivdph()\\
mDPH & mdph()\\
MPH* & MPHstar()\\
fMPH & bivph()\\
mPH & mph()\\
fMIPH & biviph()\\
mIPH & miph()\\
\bottomrule
\end{tabular}
\end{table}

\begin{table}
\caption{\label{tab:methods}Methods for matrix distributions available in matrixdist}
\centering
\resizebox{\textwidth}{!}{\begin{tabular}[!htbp]{lll}
\toprule
Method & Description & Available for\\
\midrule
sim() & Simulates iid realisations & dph, ph, iph, bivdph, mdph, MPHstar, bivph, mph, biviph, miph\\
moment() & Moments & dph, ph, bivdph, mdph, bivph, mph\\
laplace() & Laplace transform & ph, bivph, mph\\
mgf() & Moment-generating function & ph, bivph, mph\\
pgf() & Probability-generating function & dph, bivdph, mdph\\
+ & Parmetrisation of the sum of two matrix distributions & dph, ph\\
minimum() & Parmetrisation of the minimum of two matrix distributions & dph, ph, iph\\
maximum() & Parmetrisation of the maximum of two matrix distributions & dph, ph, iph\\
mixture() & Parmetrisation of a mixture of two matrix distributions & dph, ph\\
Nfold() & N-fold convolution with N DPH distributed & dph, ph\\
dens() & Density function & dph, ph, iph, bivdph, mdph, bivph, mph, biviph, miph\\
cdf() & Cumulative distribution function & dph, ph, iph, mph, miph\\
haz() & Hazard rate function & ph, iph\\
quan() & Quantiles & ph, iph\\
TVR() & Transformation via rewards & dph, ph\\
fit() & Estimation & dph, ph, iph, bivdph, mdph, MPHstar, bivph, mph, biviph, miph\\
reg() & PI regression & ph, iph\\
MoE() & MoE regression & dph, ph, iph, bivdph, mdph, mph, miph\\
marginal() & Parametrisation of marginal distribution & bivdph, mdph, MPHstar, bivph, mph, biviph, miph\\
linCom() & Parametrisation of a linear combination of marginals & MPHstar, bivph\\
mean() & Mean or vector of means & dph, ph, bivdph, mdph, MPHstar, bivph, mph\\
var() & Variance or covariance matrix & dph, ph, bivdph, mdph, MPHstar, bivph, mph\\
cor() & Correlation matrix & bivdph, mdph, MPHstar, bivph, mph\\
\bottomrule
\end{tabular}}
\end{table}


\paragraph{Data Availability Statement}
Data availability does not apply to this article as no new data were created or analysed. The \pkg{matrixdist} package is available in the Comprehensive R Archive Network (CRAN) and the GitHub repository \url{https://github.com/martinbladt/matrixdist_1.0}. The numerical examples are reproducible using the code contained in the manuscript.

\paragraph{Funding Statement}
The first author was supported by the Carlsberg Foundation, grant CF23-1096.

\paragraph{Competing Interests}
The authors declare none.

\newpage

\bibliographystyle{apalike}
\bibliography{matrixdist.bib}

\end{document}